\documentclass[a4paper,11pt]{article}
\pdfoutput=1 % if your are submitting a pdflatex (i.e. if you have
\usepackage{jheppub} % for details on the use of the package, please
\usepackage[T1]{fontenc} % if needed
\usepackage{natbib}
\usepackage{here}
\usepackage{float}
\usepackage{multirow}
\usepackage{graphicx}
\usepackage{subcaption}
\usepackage{adjustbox}
\usepackage{multirow}
\usepackage{bbm}
\usepackage{cleveref}
\usepackage{subcaption}
\usepackage{braket}

\usepackage{tikz}
\usetikzlibrary{decorations.pathmorphing}
\usetikzlibrary{decorations.markings}
\usetikzlibrary{positioning, shapes, arrows}
\usetikzlibrary{calc}
\usepackage[compat=1.0.0]{tikz-feynman}

\tikzset{
quarkar/.style={postaction={decorate}, decoration={markings,mark=at position .5 with {\arrow[#1]{latex}}},very thick},
quark/.style={postaction={decorate}, decoration={markings}, very thick},
scalar/.style={dashed,postaction={decorate}, decoration={markings,mark=at position .5 with {\arrow[#1]{latex}}}},
gluon/.style={decorate,decoration={coil,amplitude=3pt, segment length=4.7pt, pre length=.01cm, post length=.01cm}},
gluont/.style={decorate,decoration={coil,amplitude=3pt, segment length=3.50pt, pre length=.01cm, post length=.01cm}},
photon/.style={decorate, decoration={snake, segment length=6pt, amplitude=1.8pt,  pre length=1pt, post length=1pt}},
photontop/.style={/tikz/preaction={draw=white,line width=5pt}, decorate, decoration={snake, segment length=6pt, amplitude=1.8pt,  pre length=1pt, post length=1pt}},
cut/.style={red, very thick, dashed},
}

\def\nn{\nonumber}

\def\omegal{{v_2\cdot\ell }}

\usepackage[utf8]{inputenc}

\usepackage{hyperref}
\usepackage{bbold}

\usepackage{graphicx,epsf}
\usepackage{amsmath,amsfonts,amssymb,amsbsy}

\usepackage{hyperref}
\usepackage{xcolor}
\hypersetup{
    colorlinks=true,
    allcolors={blue!60!black}
}

\usepackage{multirow}
\usepackage{here}
\usepackage{stackrel}
\usepackage{tikz}
\usetikzlibrary{calc}
\usetikzlibrary{positioning} 

\usepackage{ifdraft}
\usepackage[obeyFinal]{easy-todo}
\usepackage{tensor}

\usepackage{placeins}
\usepackage{slashed} % For Dirac slashed notation

\usepackage{booktabs}
\usepackage{adjustbox}
\usepackage{bm}

\usepackage{enumerate}

\newcommand{\be}{\begin{equation}}
\newcommand{\ee}{\end{equation}}

\newcommand{\bea}{\begin{eqnarray}}
\newcommand{\eea}{\end{eqnarray}}

\newcommand{\cI}{\mathcal{I}}

\newcommand{\barm}{\bar{m}}

\def\cS{\mathcal{S}}
\def\cM{\mathcal{M}}
\def\dh{\hat{\delta}}
\def\nuv{{\vec\nu}}
\def\cIh{\widehat\cI}

\def\nn{\nonumber}

\def\spa#1.#2{\left\langle#1\,#2\right\rangle}
\def\spb#1.#2{\left[#1\,#2\right]}
\def\spash#1.#2{\spa{\smash{#1}}.{\smash{#2}}}
\def\spbsh#1.#2{\spb{\smash{#1}}.{\smash{#2}}}
\def\sand#1.#2.#3{%
\left\langle\smash{#1}{\vphantom1}^{-}\right|{#2}%
\left|\smash{#3}{\vphantom1}^{-}\right\rangle}
\def\sandpp#1.#2.#3{%
\left\langle\smash{#1}{\vphantom1}^{+}\right|{#2}%
\left|\smash{#3}{\vphantom1}^{+}\right\rangle}
\def\sandpm#1.#2.#3{%
\left\langle\smash{#1}{\vphantom1}^{+}\right|{#2}%
\left|\smash{#3}{\vphantom1}^{-}\right\rangle}
\def\sandmp#1.#2.#3{%
\left\langle\smash{#1}{\vphantom1}^{-}\right|{#2}%
\left|\smash{#3}{\vphantom1}^{+}\right\rangle}

\newcommand*{\Caravel}{\textsc{Caravel}}

\title{
Gravitational Bremsstrahlung in Black-Hole Scattering at $\mathcal{O}(G^3)$: 
Quadratic-in-Spin Effects
}

\author[1]{Lara Bohnenblust,}
\emailAdd{lara.bohnenblust@uzh.ch}
\author[1,2]{Harald Ita,}
\emailAdd{harald.ita@psi.ch}
\author[3]{Manfred Kraus,}
\emailAdd{mkraus@fisica.unam.mx}
\author[1]{Johannes Schlenk}
\emailAdd{johannes.schlenk@psi.ch}

\affiliation[1]{Department of Astrophyiscs, University of Zurich,
Winterthurerstrasse 190, 8057 Zurich, Switzerland}
\affiliation[2]{ Paul Scherrer Institut, CH-5232 Villigen PSI, Switzerland}
\affiliation[3]{Departamento de F\'{i}sica Te\'{o}rica, Instituto de
F\'{i}sica, \\ Universidad Nacional Aut\'{o}noma de M\'{e}xico, Cd. de
M\'{e}xico C.P. 04510, M\'{e}xico}

\abstract{
We are employing a supersymmetric variant of the worldline quantum field
theory (WQFT) formalism to compute the far-field momentum-space gravitational
waveform emitted during the scattering of two spinning black holes at
next-to-leading order (NLO) in the post-Minkowskian expansion. Our results are
accurate up to quadratic-in-spin contributions, which means we report for the
very first time the waveform observable at the order $\mathcal{O}(G^3\cS^2)$.
Our computation is based on mapping $n$-body tree-level amplitudes in such a
way that we can obtain the $(n-2)$-loop two-body waveform integrand. We discuss in detail
this procedure and highlight the similarity of the resulting structures with
those obtained in the scattering-amplitude approach.  As a by product of our
computational approach, we also obtain, for the first time, the leading-order
waveform for three-body scattering of spinning black holes.
We validated our results in various ways but most notably, we find exact
agreement for the NLO waveform integrand obtained from the WQFT and the
classical limit of scattering amplitudes in QFT.
}

\begin{document}
\maketitle
\newpage

%--------------------------------------------------------------------------------
    
%--------------------------------------------------------------------------------
\section{Introduction}
%--------------------------------------------------------------------------------
Since the first detection of gravitational waves~\cite{LIGOScientific:2016aoc,
LIGOScientific:2016dsl, LIGOScientific:2016sjg, LIGOScientific:2017vwq,
LIGOScientific:2017bnn} by LIGO and Virgo, the study of gravitational radiation
from compact binary systems has advanced significantly. Detector sensitivities
have steadily improved, and to date, the LIGO-Virgo-KAGRA (LVK) collaboration
has reported 90 merger events involving compact objects~\citep{KAGRA:2021vkt}.
Looking ahead, next-generation observatories such as the Einstein Telescope,
Cosmic Explorer, TianQin, and the Laser Interferometer Space Antenna (LISA)
promise to probe gravitational waves with even greater sensitivity and across a
broader frequency range~\citep{Punturo_2010,Reitze:2019iox,TianQin:2015yph,LISA:2017pwj}. These
advances will enable deeper exploration of binary parameters, including spin
and orbital eccentricity.

A central theoretical challenge in this context is the accurate and efficient
modeling of gravitational waveforms from such systems~\cite{Purrer:2019jcp,
Borhanian:2022czq}. Black holes and neutron stars are often highly spinning,
and their spins significantly affect the dynamics of the
inspiral~\cite{Wagoner:1976am, Cutler:1992tc, Blanchet:1989ki, Damour:1990ji,
Kidder:1992fr, Cutler:1994ys, Apostolatos:1994mx}. This is particularly true
when the spin vectors are misaligned with the orbital angular momentum, leading
to orbital precession and phase modulations in the gravitational waveform. 
Capturing these spin
effects is essential for building high-precision waveform models that can fully
exploit the scientific potential of current and future gravitational wave
observations.

To meet this precision challenge, several analytical frameworks have been
developed. %to tackle the relativistic two-body problem.  
Both the post-Newtonian
(PN) and, more recently, the post-Minkowskian (PM) expansions have seen
tremendous progress in the analytical study of two-body dynamics. In the PN
framework, results have been obtained through classical
methods~\cite{Krishnendu:2021cyi, Barker:1970zr, Barker:1975ae, Kidder:1992fr,
Kidder:1995zr, Tagoshi:2000zg, Faye:2006gx, Blanchet:2006gy, Damour:2007nc,
Steinhoff:2007mb, Steinhoff:2008zr, Steinhoff:2008ji, Marsat:2012fn,
Hergt:2010pa, Hergt:2012zx, Bohe:2012mr, Hartung:2013dza, Marsat:2013wwa,
Bohe:2015ana, Bini:2017pee} as well as effective field theory (EFT)
approaches~\cite{Siemonsen:2017yux, Porto:2005ac, Porto:2006bt, Porto:2007tt,
Porto:2008tb, Levi:2008nh, Porto:2008jj, Porto:2010tr, Levi:2010zu,
Porto:2010zg, Hartung:2011ea, Levi:2011eq, Porto:2012as, Levi:2014sba,
Levi:2014gsa, Levi:2015msa, Levi:2015uxa, Levi:2015ixa, Levi:2016ofk,
Maia:2017gxn, Maia:2017yok, Levi:2019kgk, Levi:2020kvb, Levi:2020uwu,
Antonelli:2020aeb, Levi:2020lfn, Antonelli:2020ybz, Goldberger:2020fot,
Liu:2021zxr, Cho:2021mqw, Kim:2021rfj,Mandal:2022nty,Mandal:2022ufb,
Bhattacharyya:2023kbh}. On the other hand, the PM expansion is
more naturally employed in quantum field theoretical methods based either on
scattering amplitudes~\cite{Cho:2022syn, Bini:2017xzy, Bini:2018ywr,
Vines:2017hyw, Vines:2018gqi, Guevara:2017csg, Guevara:2018wpp, Chung:2018kqs,
Arkani-Hamed:2019ymq, Guevara:2019fsj, Chung:2019duq, Damgaard:2019lfh,
Aoude:2020onz, Chung:2020rrz, Guevara:2020xjx, Bern:2020buy,
Kosmopoulos:2021zoq, Chen:2021kxt, FebresCordero:2022jts, Bern:2022kto,
Aoude:2022trd, Aoude:2022thd, Bern:2023ity, Menezes:2022tcs, Riva:2022fru,
Damgaard:2022jem, Bautista:2022wjf, Lindwasser:2023zwo, Gatica:2024mur,
Cristofoli:2021jas, Luna:2023uwd, Gatica:2023iws, Heissenberg:2023uvo,
Lindwasser:2023dcv, Bautista:2023sdf, Cangemi:2023ysz, Brandhuber:2024bnz,
Brandhuber:2024lgl, Alaverdian:2024spu, Chen:2024mmm, Akpinar:2024meg,
Bohnenblust:2024hkw, Bonocore:2024uxk, Bonocore:2025stf, Akpinar:2025bkt,
Bjerrum-Bohr:2025lpw, Vazquez-Holm:2025ztz} or the worldline quantum field
theory
%
% Mogull:2020sak, % original Plefka, Steinhoff et al
% Jakobsen:2021smu, PRL waveform
%Jakobsen:2021lvp, % susy WQFT for spin
%Jakobsen:2021zvh, % susy long
%Jakobsen:2022fcj, % conservative 3PM 
%Jakobsen:2022zsx, %linear Response 
%Jakobsen:2023hig, % dissipative scattering 4PM
%Jakobsen:2023ndj, % conservative 4PM
%Bhattacharyya:2024kxj, % random
%Haddad:2024ebn % high spin
%Jakobsen:2022psy % all things retarded
% Kalin:2020mvi% same, however called EFT
%Kalin:2022hph % Schwinger-Keldysh
% 
(WQFT)~%
%\cite%Mogull:2020sak, Jakobsen:2021smu, Jakobsen:2021lvp,
%Jakobsen:2021zvh, Jakobsen:2022fcj, Jakobsen:2022zsx, Jakobsen:2023hig,
%Jakobsen:2023ndj, Bhattacharyya:2024kxj, Haddad:2024ebn}
\cite{Mogull:2020sak,Jakobsen:2022psy,Kalin:2020mvi,Kalin:2022hph} to describe the
\textit{scattering} of black holes. A different and orthogonal approach is the
self-force expansion~\cite{Quinn:1996am, Mino:1996nk, Poisson:2011nh,
Barack:2018yvs, Gralla:2021qaf, Pound:2021qin, Rettegno:2023ghr, Kosmopoulos:2023bwc, Cheung:2023lnj, Cheung:2024byb, 
Driesse:2024xad, Driesse:2024feo, Akpinar:2025huz} in which one computes to all
orders in the Newton constant $G_N$ but expands in the mass-ratio of the black
holes.

Here, we focus on the PM approach and extend previous
results for black-hole scattering waveforms. The leading-order (LO) result for
non-spinning black holes has long been known from classical general
relativity~\cite{DEath:1976bbo, Kovacs:1977uw, Kovacs:1978eu}. These results
were rederived in the WQFT \cite{Jakobsen:2021smu,
Mougiakakos:2021ckm} and the up to all-order-in-spin corrections were added in
refs.~\cite{Jakobsen:2021lvp, DeAngelis:2023lvf, Aoude:2023dui,Brandhuber:2023hhl}. Since then, next-to-leading order (NLO) results
at $\mathcal{O}(G^3)$ were obtained from scattering amplitudes in a
QFT~\cite{Herderschee:2023fxh, Brandhuber:2023hhy, Georgoudis:2023lgf,
Elkhidir:2023dco,Bohnenblust:2023qmy}, using the
Kosower-Maybee-O'Connell (KMOC) formalism~\cite{Kosower:2018adc,
Maybee:2019jus, Cristofoli:2021vyo}.  In ref.~\cite{Bini:2024rsy}, consistency
between the MPM framework and amplitude-based results was established.
% too detailed for our purpose
% once
%zero-energy gravitons and disconnected diagrams in the KMOC formalism were taken
%into account. However, these additional contributions effectively amount to a
%coordinate transformation that can be obtained via BMS transformations~\cite{Elkhidir:2024izo}.  
The linear-in-spin correction to
the NLO waveform at $\mathcal{O}(G^3 \cS)$ was subsequently computed \cite{Bohnenblust:2023qmy} using amplitude methods and cross-checked
against a WQFT calculation.

In this paper, we extend the PM NLO scattering waveform by
including quadratic-in-spin corrections at ${\cal O}(G^3\cS^2)$.  To systematically organize the
classical perturbation theory, we employ both 
WQFT and standard QFT methods.  In particular, we
focus on the diagrammatic structure of the WQFT to make the mass dependence and
classical spin contributions manifest.  %, albeit with retarded propagators.
The one-loop WQFT integrand is
constructed from seven-point tree-level amplitudes, which serve as generating
building blocks (see e.g.~\cite{Shen:2018ebu,Mogull:2020sak}). In contrast to the WQFT, the amplitude-based approach offers
well-known and well-tested Feynman rules, which we use to cross-check our
results. For validation, we make use of a numerical implementation \cite{Abreu:2020xvt,Bohnenblust:2023qmy} of the
relevant amplitudes.  This comparison also includes an implementation of the
Casimir $\cS^2$ contributions, which is based on the \textit{spin-interpolation}
method of refs.~\cite{Akpinar:2024meg,Akpinar:2025bkt} and confirms its
validity. The spectral waveform requires momentum integration associated to the exchanged radiation, which resembles
the one of one-loop Feynman integrals.  
The retarded momentum integrals are shown to match the known integral basis 
\cite{Caron-Huot:2023vxl} obtained within the KMOC formalism.

This article is organized as follows: In section~\ref{sec:notation} we state
our conventions and introduce our choice of kinematical variables. In
section~\ref{sec:wqft} we briefly summarize the WQFT formalism and lay out our
strategy to compute the gravitational waveform at NLO.  Then, in
section~\ref{sec:computation} we give further details on the explicit
computation. In section~\ref{sec:QFT} we discuss how to perform the
corresponding calculation in a scattering-amplitude setup. In
section~\ref{sec:results} we present our results and in
section~\ref{sec:conclusions} we summarize our conclusions.

%--------------------------------------------------------------------------------
\section{Setup, notation and kinematics}
\label{sec:notation}
\begin{figure}[ht]
  \centering
  \includegraphics[width=0.5\linewidth]{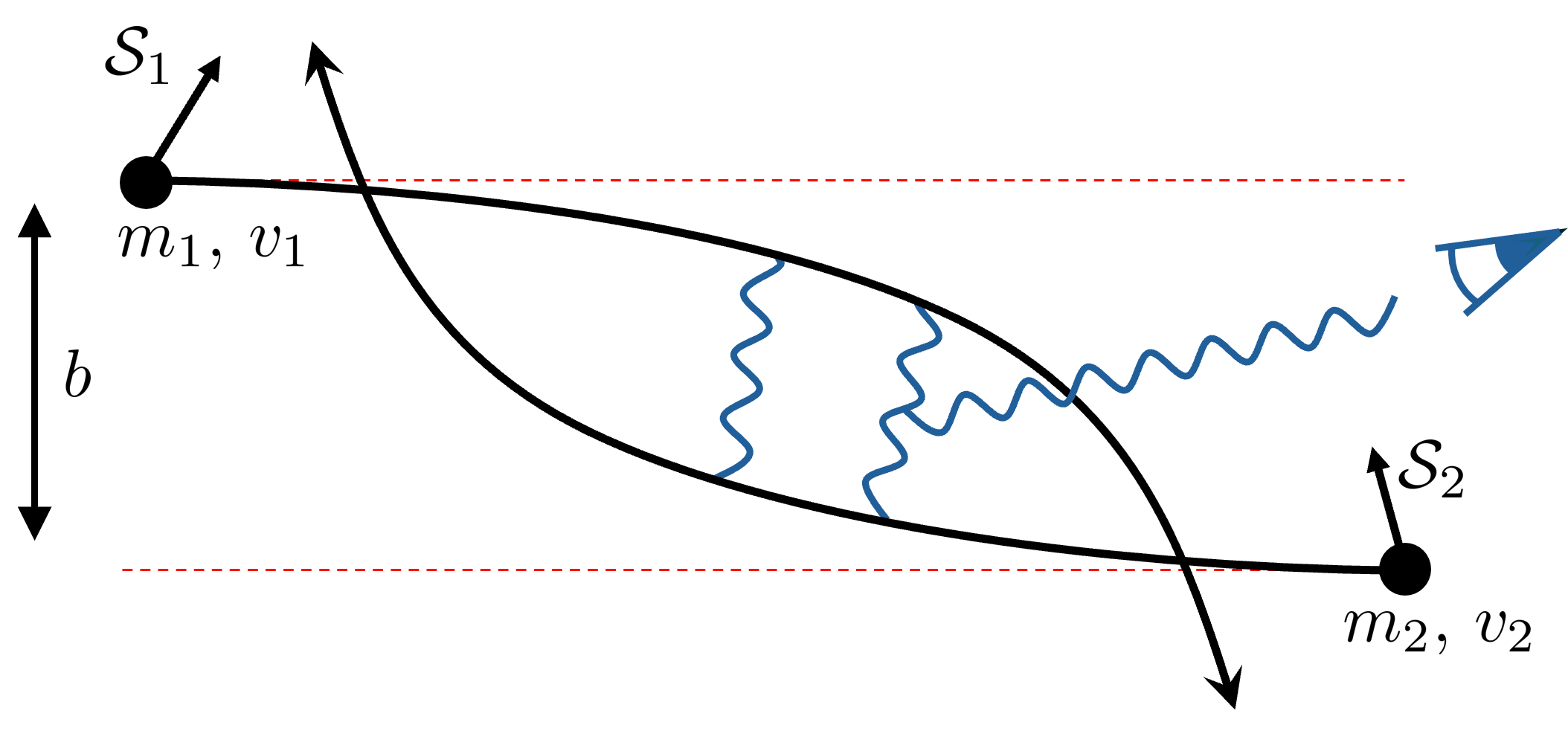}
  \caption{The scattering process of two massive compact objects that emits
  gravitational radiation. The scattering is mediated by the exchange of
  gravitational waves shown in blue. The initial masses, velocities and spins
  are indicated.}
  \label{fig:process}
\end{figure}

%--------------------------------------------------------------------------------
We consider the scattering of two compact
spinning objects, described as point masses ($m_i$) and with spin $\cS_i$. 
We assume the objects Compton wavelength $\lambda_c\sim 1/m_i$ to be much smaller 
than their Schwarzschild radius $r_{s,i}=2 G m_i$, such 
that $G m_i^2 \gg 1$. % and the objects long-range gravitational field is classical. 
The scattering is assumed to take place with an impact parameter $b=|b_1-b_2|$ much
larger than the objects' Schwarzschild radius, which implies the
classical weak-field regime with $b\gg r_{s,i}$. The spin is assumed to be
classical and small enough, such that the ring radius $a_i=\cS_i/m_i$ is much smaller 
than the Schwarzschild radius  $a_i/r_{s,i}\ll 1$, and we treat it perturbatively.  
The system emits gravitational radiation and we compute its waveform
at large distance $|r|\gg b$ from the scattering objects at 
$\mathcal{O}(G^3 \cS^2)$. A pictogram of the considered process with 
relevant quantities indicated is shown in \cref{fig:process}.
Below we will mostly refer to the scattering objects as black holes for simplicity.

In the remainder of this section, we set up our notation and conventions for the
kinematic variables which are used to express all obtained scattering observables.
We are working in the mostly-minus metric signature, such that the asymptotic 
flat-space metric is given by 
\begin{equation}
\eta_{\mu \nu} = \eta^{\mu \nu} = \mathrm{diag}(1,-1,-1,-1)\;.
\end{equation}
We consider a generic scattering process of $n$ compact objects, such as black holes, 
which produces Bremsstrahlung with momentum $k$,
\begin{equation}
 (m_1 v_1, \ldots, m_n v_n) \quad\rightarrow\quad (m_1 v_1-q_1, \ldots, m_n v_n-q_n, k)\,,
\end{equation}
where we associate a mass $m_i$, a velocity $v_i$ and a momentum exchange $q_i$
to each black hole. The velocities are normalized as $v_i^2 = 1$ and momentum
conservation relates the momentum of the emitted gravitational radiation $k$ to the exchanged
momenta $q_i$ via
\begin{align}
k=\sum_{i=1}^n q_i\,.
\end{align}
Next, we define kinematic invariants
that parameterize the aforementioned scattering processes, following the
conventions in ref.~\cite{Mogull:2020sak},
\begin{align}
\omega_i = v_i\cdot k\;, \quad 
\omega_{ij} = v_i \cdot q_j\;, \quad
q_i^2\;, \quad 
q_{ij}^2 = (q_i+q_j)^2\;, \quad
y_{ij} = v_i\cdot v_j\;,
\end{align}
which are subject to the following constraints
\begin{align}
\omega_i = \sum_{j=1}^n \omega_{ij}\;, \quad
\omega_{ii}=0\;,\quad
k^2 = \left(\sum_i q_i\right)^2=\sum_{i< j}^n q_{ij}^2-(n-2)\sum_{i=1}^n q_i^2=0\;.
\end{align}
The classical amplitudes in the PM expansion are polynomial in the masses $m_i$ of 
the compact objects and one can obtain their dependence directly from diagrammatic 
rules, as will be discussed below. Consequently, we can omit the $m_i$ parameters 
in most computations.
We do not use Gram-determinant identities to eliminate momentum redundancies
which originate from the four dimensionality of spacetime, instead we assume
general $d$-dimensional external momenta.

In this paper, we focus on two cases: two-body scattering ($n=2$) and
three-body scattering ($n=3$).  
For the binary system there are only $5$ variables,
\begin{equation}
 \omega_1\;, \quad \omega_2\;, \quad q_1^2\;, \quad q_2^2\;, \quad y \equiv y_{12}\;.
 \label{eq:2BodyVariables}
\end{equation}
For three-body scattering, we
write our results in terms of the following $14$ variables
\begin{equation}
\begin{split}
 &\omega_{ij}\quad\mbox{for}\quad i,j=1,2,3\quad \mbox{and}\quad  i \neq j \;, \\
 &y_{ij}\,,\,\, q_{ij}\quad\mbox{for}\quad i,j=1,2,3\quad \mbox{and}\quad  i<j \;, \\
 &q_1^2\quad \mbox{and}\quad  q_2^2 \;.
\label{eq:3BodyVariables} 
\end{split}
\end{equation}
In position space, the sources' initial trajectories will be the straight worldlines,
\begin{align}
x_i^\mu(\tau) = b_i^\mu + \tau \, v_i^\mu \,,
\end{align}
which depend on the velocities $v_i$ and the impact parameters $b_i$. Here
$\tau$ is the proper time of the moving object.  Finally, the objects'
initial spin will be denoted by a rank-two tensor, $\cS_i^{ab}$, which in
turn is related to the classical spin vector
\begin{equation}
  \cS_i^{a} = \frac{1}{2}\epsilon^a_{~bcd} v_i^{b} \cS_i^{cd}  \,,
\end{equation}
with the convention $\epsilon^{0123}=1$. We want to stress that the spin
vectors $\cS_i^a$ in the WQFT are defined in their respective inertial
frames~\cite{Jakobsen:2022zsx} corresponding to $v_i^b$. This choice of
reference frame matches the commonly chosen frame in the scattering-amplitude
approach. Throughout our computation we impose the covariant spin-supplementary 
condition,
\begin{equation}\label{eqn:ssc}
  \cS_i^{ab} v_{ia} = 0  \,,
\end{equation}
to uniquely specify the spin variables 
\cite{Jakobsen:2022zsx,Bautista:2023szu}.
Finally, it will often be useful to group a mass factor with the associated spin vector, 
\begin{equation}\label{eqn:ringRadius}
  a_i^{c} = \frac{\cS_i^{c}}{m_i}  \,,
\end{equation}
to emphasize the parameter dependences. The vectors $a_i$ have the dimension of length and 
its norm $|a_i|$ corresponds to the ring radius of the rotating black hole.

For later convenience, we also introduce the following integral measures
\begin{equation}
  \int_\omega \equiv \int_{-\infty}^\infty \frac{d\omega}{2\pi}\;, \qquad 
  \int_k \equiv \int \frac{d^dk}{(2\pi)^d}\;,
\end{equation}
and the shorthand,
\begin{equation}
  \hat\delta(x) = 2 \pi \delta ( x) \,.
\end{equation}
At last, we define the $d$-dimensional physical-state projector
$P_{\mu\nu\rho\sigma}$ for the gravitational field as
\begin{equation}
    P_{\mu\nu\rho\sigma} = \frac{1}{2}\left[\eta_{\mu\rho}\eta_{\sigma\nu} + \eta_{\mu\sigma}\eta_{\rho\nu} 
   -\frac{2}{d-2}\eta_{\mu \nu}\eta_{\rho\sigma}\right]\;.
\end{equation}

%--------------------------------------------------------------------------------
\section{Waveforms from the worldline QFT formalism}
\label{sec:wqft}
%--------------------------------------------------------------------------------
We study the radiation emitted during the encounter of two spinning black holes
by using the worldline quantum field theory (WQFT)
approach~\cite{Jakobsen:2022psy,Mogull:2020sak,Kalin:2020mvi,Kalin:2022hph}. The formalism has been extended \cite{Jakobsen:2021lvp,Jakobsen:2021zvh} 
to obtain the leading-order waveforms up to quadratic spin effects
\cite{Jakobsen:2021smu,Jakobsen:2021lvp}, and was recently further generalized
\cite{Haddad:2024ebn} to include higher-spin contributions. However, in this
work we use a simplified supersymmetric variant~\cite{Jakobsen:2021lvp,Jakobsen:2021zvh}
suitable for obtaining spin-squared effects.  In the following section, we
review the main ideas of the WQFT approach for spinning sources. 
In particular, we provide a detailed elaboration on extracting the classical gravitational waveform in terms of many-body scattering diagrams.

%--------------------------------------------------------------------------------
\subsection{The worldline theory and weak-field expansion}
%--------------------------------------------------------------------------------
To describe black-hole scattering, we let $N$ massive spinning bodies interact
gravitationally 
\begin{align}
  S = S_{\mathrm{EH}} + S_{\mathrm{gf}} + \sum_{i=1}^N S_i\;,
\label{eq:FullAction}
\end{align}
where the Einstein-Hilbert action $S_{\mathrm{EH}}$ is given by
\begin{equation}
  S_{\mathrm{EH}} = -\frac{2}{\kappa^2}\int d^4 x \sqrt{|g|}R\,,
\label{eq:EHaction}
\end{equation}
with $g=\det (g_{\mu\nu})$ and the Ricci scalar is defined by
$R=R^{\rho}_{~\mu\rho\nu}\,g^{\mu\nu}$.  We will assume weak gravitational
fields, such that one can expand the spacetime metric in a fluctuation
$h_{\mu\nu}$ around flat space via
\begin{equation}\label{eq:metricFluctuation}
  g_{\mu\nu}(x) \equiv \eta_{\mu\nu}+\kappa\; h_{\mu\nu}(x)\;,
\end{equation}
where $\kappa = \sqrt{32 \pi G}$, which is written in terms of the Newton
constant $G_N$ using $G=\hbar G_N/c^3$.  
The gauge fixing-term $S_{\mathrm{gf}}=\int d^4x\,
\eta_{\mu\nu} f^\mu f^\nu$ is used to impose the de Donder gauge, $f^\nu=\partial_\mu
h^{\mu\nu}-\frac{1}{2}\partial^\nu h^\mu_{~\mu} =0$.  Up to quadratic order in
spin, the dynamics of a spinning massive body can be described by the worldline
action \cite{Jakobsen:2021zvh}
\begin{equation}\label{eq:WLaction}
  S_i = -m_i \int d\tau \left[ \frac{1}{2}g_{\mu\nu}\dot{x}_i^{\mu}\dot{x}_i^{\nu} 
  + i \bar{\psi}_{i,a} \frac{D\psi_i^a}{D\tau} 
  + \frac{1}{2}R_{a b c d} \bar{\psi}_i^a \psi_i^b \bar{\psi}_i^c \psi_i^d \right]\,,
\end{equation}
where $\dot{x}^\mu \equiv \partial x^{\mu}/\partial \tau$ and the covariant derivative is
defined via $\frac{D\psi_i^a}{D\tau} = \dot{\psi}_i^a + \dot{x}^\mu
\omega_\mu^{~ab}\psi_{i,b}$ with the spin connection $\omega_\mu^{~ab}$.  To
treat the fermionic degrees of freedom we have introduced the Vielbein
$e_{\mu}^a(x)$, which is defined through the relation
\begin{equation}\label{eq:VielbeinRelation}
  g_{\mu\nu}(x) = e_{\mu}^a(x)\; e_{\nu}^b(x)\; \eta_{a b}\;.
\end{equation}
Here the Roman letters are associated to the orthonormal tangent space related
to the curved spacetime denoted by Greek indices. In the asymptotic flat
spacetime, where we assume tangent and coordinate spaces to be aligned, we stop
distinguishing Greek and Roman indices.  The spin connection is linked to the
Christoffel symbols by
\begin{equation}
    \omega_\mu^{~ab} \equiv e^a_\nu \left( \partial_\mu e^{\nu b} + \Gamma^\nu_{\mu\lambda} e^{\lambda b}\right)\;.
\end{equation}
The dynamical variables are the trajectories $x^\mu_i(\tau)$ and the Grassmann
worldline fields $\psi^a(\tau)$ and $\bar{\psi}^a(\tau)$. The interaction term
including the Riemann tensor requires the Vielbein fields (\ref{eq:VielbeinRelation}) 
to relate spacetime- to tangent-space indices.

To describe the scattering of black holes, we expand the trajectories around
their initial configuration and wish to obtain the back reaction to the metric
and the paths in perturbation theory.  We therefore parameterize the paths via
\begin{align}
    x_i^{\mu}(\tau) = b_i^{\mu} + v_i^{\mu} \tau + z_i^{\mu}(\tau)\;,
\end{align}
where $b_i$ is the impact parameter of particle $i$ and $v_i$ its initial
velocity.  (The initial momentum of the massive object is $p_j=m_j v_j$.)
Finally, $z_i$ represents the perturbation around the background trajectory and
is a new dynamical field, for which we impose the initial condition
$z_i(-\infty)=0$.  Furthermore, the Grassmann fields are also expanded around
the background
\begin{equation}
 \psi_i^a(\tau) = \Psi_i^a + \phi_i^a(\tau)\,,
\end{equation}
with the perturbation $\phi_i^a(\tau)$. We require the field fluctuations to
vanish in the infinite past, i.e.  $\phi_i^a(-\infty)=0$.  The constant part,
$\Psi_i^a$ is related to the classical spin tensor~\cite{Jakobsen:2021zvh}
\begin{equation}
  \mathcal{S}_i^{a b} = -2i m_i \bar{\Psi}_i^{[a}\Psi_i^{b]}\;,
\end{equation}
where we used the following (anti-)symmetrization
\begin{equation}\label{eqn:tensorSym}
	x^{[a}y^{b]} \equiv \frac{1}{2}\left(x^a y^b - x^b y^a\right)
	\quad\mbox{and}\quad
	x^{(a}y^{b)} \equiv \frac{1}{2}\left(x^a y^b + x^b y^a\right)\;.
\end{equation}
The parameters $b_i$, $v_i$ and $\cS_i$ will set the initial conditions of a
scattering process.

Working in the perturbative weak-field regime, the inverse metric $g^{\mu\nu}$
and the determinant of the metric are expanded in $\kappa$,
\begin{align}
  g^{\mu\nu} &= \eta^{\mu\nu} - \kappa h^{\mu\nu} + \kappa^2 h^{\mu\lambda}h^\nu_{~\lambda} - \kappa^3 h^{\mu\lambda}h_{\lambda\sigma}h^{\sigma\nu} 
  + \mathcal{O}(\kappa^4)\;, \\
  \sqrt{|g|} &= 1 + \frac{\kappa}{2}h + \frac{\kappa^2}{8}\Big(h^2 - 2h_{\mu\nu}h^{\mu\nu}\Big) \nonumber \\
  &+ \frac{\kappa^3}{48}\Big(h^3 - 6hh^{\mu\nu}h_{\mu\nu} + 8 h^{\mu\nu}h_{\nu\lambda}h^\lambda_{~\mu}\Big) + \mathcal{O}(\kappa^4)\;,
\end{align}
where $h\equiv h^{\mu}_{~\mu}$.  Furthermore, we need the expansion of the
covariant derivative $\nabla_\mu$, for which the required Christoffel symbols
$\Gamma^\lambda_{\mu\nu}$ are expanded as
\begin{equation}
  \Gamma^\lambda_{\mu\nu} = \frac{\kappa}{2}\left(\eta^{\lambda\omega} - \kappa h^{\lambda\omega} + \kappa^2 h^{\lambda\sigma}h_{\sigma}^{~\omega}\right)
  \left(\partial_\mu h_{\omega\nu} + \partial_\nu h_{\mu\omega} - \partial_\omega h_{\mu\nu}\right) + \mathcal{O}(\kappa^4)\;.
\end{equation}
To treat spin, we must also expand the Vielbein $e^a_\mu$ and the
spin-connection $\omega_{\mu}^{~ab}$, which up to third order read
\begin{align} \label{eq:Vielbein}
  e^a_\mu &= \eta^{a \nu}\left(\eta_{\mu\nu} + \frac{\kappa}{2}h_{\mu\nu} - \frac{\kappa^2}{8} h_{\mu\rho}h^{\rho}_{~\nu} 
  + \frac{\kappa^3}{16}h_{\mu\rho}h^{\rho}_{~\sigma}h^\sigma_{~\nu}\right) + \mathcal{O}(\kappa^4)\;,\\
  \omega_{\mu}^{~ab} &= -\kappa \partial^{[a}h^{b]}_{~\mu}
  - \frac{\kappa^2}{2}h^{\nu [a}\left(\partial^{b]}h_{\mu\nu} - \partial_{\nu}h^{b]}_{~\mu}+\frac{1}{2}\partial_{\mu}h^{b]}_{~\nu}\right) \nonumber \\
  &-\frac{\kappa^3}{8} h^{\nu[a}\left(2 h^{b] \rho} \partial_{\nu}h_{\mu \rho} - 2 h_\nu^{~\rho}\partial_{\mu}h^{b]}_{~\rho} 
  + 3 h_\nu^{~\rho}\partial_{\rho}h^{b]}_{~\mu} - 3 h_\nu^{~\rho}\partial^{b ]}h_{\mu \rho} \right) + \mathcal{O}(\kappa^4)\;.
\end{align}
Finally, the inverse Vielbein obtains the expansion
\begin{equation}
 e_{a}^\mu =  \eta^{\mu\nu}\left(\eta_{a \nu} - \frac{\kappa}{2}h_{a\nu} + \frac{3\kappa^2}{8} h_{\nu\rho}h^\rho_{~a} 
 - \frac{5\kappa^3}{16}h_{\nu\rho}h^\rho_{~\sigma}h^\sigma_{~a}\right) + \mathcal{O}(\kappa^4)\;.
\end{equation}

%--------------------------------------------------------------------------------
\subsection{The gravitational waveform}
%--------------------------------------------------------------------------------
Having defined the dynamics of the system, we now discuss the gravitational-wave
observable which we wish to compute.  
During a scattering event, the two black holes accelerate and emit gravitational radiation 
which is measured in observatories. We want to compute the radiation's waveform
at large distance $|\mathbf{r}|\gg b$ from the scattering objects. 
The radiated metric perturbation $\kappa h^{\mu \nu}$ of two scattering black
holes is obtained by solving the classical Einstein field equations
perturbatively in the Newton coupling $G_N$ (or equivalently $\kappa$).  The
waveform $h_s^\infty$ is then identified as the coefficient of the leading
contribution in the $1/|\mathbf{r}|$ expansion of the metric fluctuation,
\begin{align}
   \kappa\, \varepsilon_s^{\mu\nu}h^{\rm cl}_{\mu\nu}(r;\cS,p,b)\Big|_{|\mathbf{r}|\rightarrow\infty}=
\frac{1}{|{\bf r}|} h_s^{\infty}(t-|{\bf r}|,{\bf n};\cS,p,b) + {\cal O}(|{\bf r}|^0) \,, 
\end{align}
where $k=\omega \hat k=\omega (1,{\bf n})$, with ${\bf n}=\bf{r}/|\bf{r}|$. The
label $s\in\{\times,+\}$ specifies the polarization $\varepsilon_s^{\mu\nu}$ of
the radiation, and the parameters $\cS$, $p$ and $b$ specify the objects' spin,
their initial momentum and impact parameter, respectively.  The time-domain waveform
$h_s^{\infty}(t-|{\bf r}|,{\bf n};\cS,p,b)$ is related to the momentum-space
waveform by the Fourier transform, 
\begin{align} \label{eq:DefinitionWaveform}
h_s^{\infty}(t-|{\bf r}|,{\bf n};\cS,p,b) = \frac{\kappa}{4\pi} \int_0^\infty
\frac{d\omega}{2\pi} e^{-i\omega(t-|{\bf r}|)}
\left[k^2\, \varepsilon_s^{\mu\nu}(k)\,h^{\rm cl}_{\mu\nu} (k;\cS,p,b) \right]_{k=\omega \hat
k} + c.c. \,,
\end{align} 
of the on-shell momentum-space metric fluctuation $h_{\mu\nu}(k; \cS,p,b)$. (To
simplify notation, we will often suppress the initial-data and denote it by
$h_{\mu\nu} (k)$.)  Consequently, the central quantity we need to obtain is the
on-shell momentum-space waveform,
\begin{align} \label{eq:waveformMomSpace}
O=k^2\,\varepsilon_s^{\mu\nu}(k)\,h^{\rm cl}_{\mu\nu}(k;\cS,p,b)
 \Big|_{k=\omega \hat k}\,.
\end{align}
We will compute the quantity \eqref{eq:waveformMomSpace} in a perturbative
expansion, 
\begin{align} \label{eq:pertExpansion}
O= \sum_{i,j,k,l\ge 0} O^{ijkl}
\left( \frac{r_{s,1}}{b} \right)^i
\left( \frac{r_{s,2}}{b} \right)^j
\left( \frac{\cS_1/m_1}{b} \right)^k
\left( \frac{\cS_2/m_2}{b} \right)^l \,,
\end{align}
where the expansion \eqref{eq:pertExpansion} is in 
terms of the ring radius $a_i=S_i/m_i$ \eqref{eqn:ringRadius}.
We will focus on contributions at next-to-leading order in the
gravitational coupling and up to second order in spin, i.e. $\mathcal{O}(G^3
\cS^2)$ using the $G$-dependence of $r_{s,i}\sim G$ and the spin $\cS$ as 
convenient counting parameters. 

%--------------------------------------------------------------------------------
\subsection{Classical metric from worldline QFT}
\label{sec:in-in}
%--------------------------------------------------------------------------------
The classical gravitational Bremsstrahlung is emitted as a non-trivial metric
perturbation, which we compute perturbatively in the gravitational coupling $G$ and spin $\cS$.  We
choose to work in a worldline QFT setup \cite{Jakobsen:2022psy, Kalin:2022hph}
to obtain a diagrammatic approach based on modern field-theory methods.  This
approach was already successfully applied to the LO $\mathcal{O}(G^2
\cS^2)$ waveforms including spin corrections \cite{Mogull:2020sak,
Jakobsen:2021zvh}, and we will focus on the NLO corrections
at $\mathcal{O}(G^3)$. Alternative field-theory approaches have provided the
waveform observable at LO at $\mathcal{O}(G^2)$ including all-order
in spin corrections \cite{Brandhuber:2023hhl, DeAngelis:2023lvf, Aoude:2023dui,
Brandhuber:2024qdn}, and at NLO at $\mathcal{O}(G^3)$
\cite{Herderschee:2023fxh, Brandhuber:2023hhy, Georgoudis:2023lgf,
Elkhidir:2023dco, Caron-Huot:2023vxl, Bohnenblust:2023qmy} including
linear-in-spin contributions \cite{Bohnenblust:2023qmy}.  We will apply the
latter type of approach for validation for which further details can be found in
\cref{sec:QFT}.

In the worldline QFT approach the metric fluctuation is determined from the
"in-in" one-point function $\langle h_{\mu \nu}(k)\big\rangle_{\rm in-in}$ of
the Keldysh-Schwinger path integral \cite{Schwinger:1960qe,Keldysh:1964ud}.
The classical field $h^{\rm cl}_{\mu \nu}$ is obtained by omitting quantum
corrections,
\begin{align}
h^{\rm cl}_{\mu \nu}(k)&= \big\langle h_{\mu \nu}(k)\big\rangle_{\rm in-in}\Big|_{\rm classical}\,.
\end{align}
In the following, we will closely follow the path-integral formulation of
ref.~\cite{Jakobsen:2022psy}, which we refer to for details on the formalism.
Here, we collect the key points of the method:
\begin{enumerate}
\item In the classical limit, the expectation value $\langle h_{\mu
\nu}(k)\big\rangle_{\rm in-in}$ is obtained from connected tree-level diagrams
of the worldline theory coupled to gravity \eqref{eq:FullAction}.
\item As opposed to a standard Feynman-diagram computation, the adjustment for
computing the "in-in" one-point function amounts to using retarded propagators
instead of Feynman propagators.
\item The formalism provides a mechanism to assign the retarded propagators
matching the causality of the process.  The latter aspect is simple to account
for in tree-level diagrams, which have a well defined causality flow from sources to
the final emitted gravitational radiation.
\item The relevant vertex rules for the Keldysh-Schwinger path integral are
equivalent to the vertex rules derived from the worldline
\eqref{eq:WLaction} 
and Einstein-Hilbert action \eqref{eq:EHaction}
\cite{Jakobsen:2022psy}.
\end{enumerate}
We thus have available a simple diagrammatic approach to obtain the classical
metric fluctuation $h^{\rm cl}_{\mu\nu}(k)$.
%--------------------------------------------------------------------------------
\subsection{Feynman rules}
%--------------------------------------------------------------------------------
In the WQFT approach we promote the perturbation fields $z_i^\mu(\tau)$,
$\phi_i^a(\tau)$ and $h_{\mu\nu}(x)$ to quantum fields. The fields are used
in their momentum-space variants,
\begin{align}
 z_i^{\mu}(\tau) = \int_{\omega} e^{i\omega \tau} z_i^{\mu}(-\omega)\;,\quad 
 \phi_i^{\mu}(\tau) = \int_{\omega}e^{i\omega\tau}\phi_i^{\mu}(-\omega)\;,\quad 
 h_{\mu\nu}(x) = \int_{k}e^{i k\cdot x}h_{\mu\nu}(-k)\;,
\end{align}
which allows us to derive Feynman rules. Given that in the worldline action
$S_i$ the graviton field implicitly depends on the trajectory, it must be
expanded using
\begin{equation}
\begin{split}
  h_{\mu\nu}(x_i(\tau)) &= \int_k e^{i k\cdot (b_i + v_i \tau + z_i(\tau))} h_{\mu\nu}(-k) \\
  &= \sum_{n=0}^{\infty} \frac{i^n}{n!}\int_k e^{i k\cdot (b_i + v_i \tau)}\big[k\cdot z_i(\tau)\big]^n h_{\mu\nu}(-k)\\
  &= \sum_{n=0}^{\infty} \frac{i^n}{n!}\int_{k,\omega_1,...,\omega_n} e^{i k\cdot b_i} e^{i (k\cdot v_i 
	+\sum_{j=1}^{n}\omega_j)\tau}\left(\prod_{j=1}^n k\cdot z_i(-\omega_j)\right) h_{\mu\nu}(-k)\;,
\end{split}
\end{equation}
which leads to polynomial interaction terms.  Due to this dependence on the trajectory,
we will obtain vertices with an arbitrary
numbers of $z$-fields from \cref{eq:WLaction}.  In contrast, the number of $\phi$-fields in a given
vertex is constrained to be at most four due to the third term in the worldline
action \eqref{eq:WLaction}. Furthermore, plugging the above
expressions into the action allows to integrate out the proper time $\tau$,
which gives rise to energy-conserving delta functions $\delta(k\cdot v_i
+\sum_{j=1}^{n}\omega_j)$.  The Feynman rules can now be extracted by
considering the path integral over the fluctuating fields with the action of
\eqref{eq:FullAction} and taking functional derivatives. 
For details on the Feynman vertices we refer the reader to
refs.~\cite{Mogull:2020sak, Jakobsen:2021zvh} and
appendix~\ref{app:feynmanrules}.

We conclude this section by collecting the required propagator conventions. The
retarded momentum-space graviton propagator is given by
\begin{equation}
 \includegraphics[]{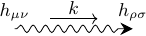}  \quad = 
 \quad i \frac{P_{\mu\nu\rho\sigma}}{(k^0+i \epsilon)^2-\bm{k}^2} = i \frac{P_{\mu\nu\rho\sigma}}{k^2 +i \epsilon \, \mathrm{sgn}(k^0)}\;,
 \label{eq:gravitonProp}
\end{equation}
with
\begin{equation}
 \mathrm{sgn}(x) \equiv \begin{cases} 1 & x > 0 \\ 0 & x = 0 \\ -1 & x < 0\end{cases}\;.
\end{equation}
The propagator for the $z$-field is
\begin{equation}
    \raisebox{0pt}{\includegraphics[]{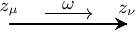} }\quad = \quad -i \frac{\eta_{\mu\nu}}{m_i(\omega+i \epsilon)^2}\,,\label{eq:zProp}
\end{equation}
and the propagators for the $\phi$-field and $\bar{\phi}$-field are
\begin{equation}
 \raisebox{0pt}{\includegraphics[]{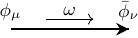} }\,\, = \,\,  \frac{-i\,\eta_{\mu\nu}}{m_i(\omega+i \epsilon)}\,,\quad\quad 
 \raisebox{0pt}{\includegraphics[]{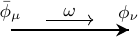} }\,\, = \,\,  \frac{-i\,\eta_{\mu\nu}}{m_i(\omega+i \epsilon)}\,.
 \label{eq:phiProp}
\end{equation}
In the diagrammatic representation of the propagators we indicate the momentum
(frequency) definition with an additional arrow, while the arrow on the diagram
line itself specifies the causality flow, i.e. if momentum arrows are aligned
the signs of $k^0$ ($\omega$) and $i\epsilon$ match in
eqns.~\eqref{eq:gravitonProp}, \eqref{eq:zProp}, and \eqref{eq:phiProp}. We
suppress the conventional arrow for fermion lines, since both propagators are equivalent.

To fix our conventions we also give the position-space Green functions for
the metric fluctuation,
\begin{align}
\includegraphics[]{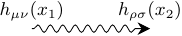}  \quad = 
\quad \int_k \frac{i\, P_{\mu\nu\rho\sigma}}{(k^0+i \epsilon)^2-\bm{k}^2} e^{-i k\cdot (x_2-x_1)} \,,
\end{align}
and the fields $z(\tau)$,
\begin{align}
\includegraphics[]{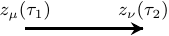}  \quad = 
\quad \int_\omega \frac{-i\,\eta_{\mu\nu}}{m_i (\omega+i \epsilon)^2}  e^{-i \omega (\tau_2-\tau_1)} \,,  
\end{align}
and similar for the Grassmann fields. The Green functions have non-vanishing (retarded)
support for $x^0_2>x^0_1$ and $\tau_2>\tau_1$, respectively.

%--------------------------------------------------------------------------------
\subsection{Classical amplitudes}
\begin{figure}[ht!]
 \centering 
 \begin{subfigure}[t]{0.49\textwidth} 
   \centering
   \includegraphics[width=0.8\textwidth]{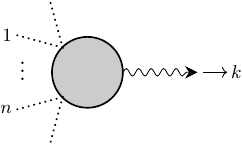} 
   \caption{}
   \label{fig:waveform} 
 \end{subfigure} 
 \begin{subfigure}[t]{0.49\textwidth}
   \centering 
   \includegraphics[width=0.8\textwidth]{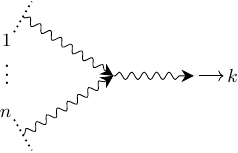} 
   \caption{}
  \label{fig:waveformEx} 
\end{subfigure} 
\caption{Left: Diagrammatic representation of the required amplitude for the
gravitational waveform. Right: An example diagram contributing to the
amplitude. Dotted lines denote background fields $\Psi_i^a$, while wavy lines
denote graviton field $h_{\mu\nu}$.}
\end{figure}

%--------------------------------------------------------------------------------
The one-point function $\langle h^{\mu\nu} \rangle_{\mathrm{in-in}}$ has a
diagrammatic expansion in terms of WQFT Feynman diagrams, which we now analyse
to simplify their computation.  Its schematic form is depicted in
\cref{fig:waveform}, where a number of worldlines (dotted lines) interact via
graviton exchange to produce the metric fluctuation $h^{\rm cl}_{\mu\nu}$.  The
external graviton leg represents a graviton-field propagator.  The worldlines support
the dynamical fields $z_i$ and the fermionic fields $\phi_i^a$ which propagate
and interact.  The propagating worldline fields are displayed by solid lines
(see e.g. diagrams \cref{fig:3BodyDiagrams}).  The interaction vertices may
include the background data $v_i$ and $\Psi_i^a$ as well as the masses $m_i$.
In \cref{fig:waveformEx}, we show an exemplary n-body contribution to the
expectation value at the leading order in perturbation theory.  Here n
worldlines emit a graviton field each, which produce the final classical metric
fluctuation with momentum $k$.  The diagram contributes at order
$\kappa^{2n-1}$ and is proportional to the mass monomial $(m_1\cdots m_n)$,
which is manifest from the $\kappa m_i$ dependence of the graviton-worldline
vertex \eqref{eq:h} and the $\kappa^{n-1}$ dependence of the (n+1)-point
graviton vertex. Diagrams may have additional mass dependence linked to spin vectors only; 
explicitly this dependence is of the form $a_i=(S_i/m_i)$, as again manifest in the Feynman rules 
(see \cref{app:feynmanrules}). Below, we find it convenient to write diagrams in terms of
the ring radius $a_i$. In this way, the diagrams have a homogeneous scaling in the mass parameters.

In a waveform computation, we do not require the off-shell metric field
$h^{\rm cl}_{\mu\nu}$ itself, since a number of simplifications appear in expression
\eqref{eq:waveformMomSpace}: First of all, the waveform computation requires to
multiply the classical field with the inverse propagator $k^2$.  Furthermore,
we require the classical field contracted with the polarization state
$\varepsilon^{\mu\nu}_s$. Finally, it is sufficient to consider the on-shell
($k^2=0$) graviton-field  momentum. We refer to this amputated one-point function as
the scattering amplitude in impact-parameter space and we use the notation,
\begin{align}
\widehat {\cal M}_s \equiv \kappa\, k^2 \varepsilon_s^{\mu\nu} h^{cl}_{\mu\nu}(k)\,.
\end{align}
The object contains an infinite sum of WQFT diagrams, which we now wish to organize, 
so that we can tackle them.
%The superscript $n$ indicates that the amplitude corresponds to the n-body
%process.  

The diagrammatic organisation of the WQFT approach makes manifest a number of
useful properties \cite{Mogull:2020sak,Jakobsen:2021lvp,Jakobsen:2021zvh,Jakobsen:2022psy,Driesse:2024xad}.  
To exploit these, it is useful to introduce the notion of a
generating amplitude, which is the contribution to $\widehat {\cal M}_s$ from
precisely n background worldlines, 
\begin{align}\label{eqn:genAmpl}
\widehat {\cal M}_s^n(1,2,\ldots, n) = 
\widehat {\cal M}_s  \Big|_{\rm n \,\,sources}\,.
\end{align}
% 
%The superscript $n$ indicates that in $\widehat {\cal M}_s^n$ only the diagrams of $\widehat {\cal M}_s$ 
%with precisely $n$ background worldlines
%are kept. 
The integer arguments specify the type of participating
worldlines, and refer to the subscript of the worldline actions $S_i$ of
\cref{eq:WLaction}.  We now discuss the properties of this generating object in
more detail. 

%--------------------------------------------------------------------------------
\paragraph{Mass and coupling dependence:}
%--------------------------------------------------------------------------------
The generating amplitudes have a specific mass-scaling and perturbative order,
\begin{align}
\widehat{\cal M}_s^{n}(1,2,3\ldots, n) \sim G^n\, \prod_{i=1}^n m_i \,.
\end{align}
For this scaling to hold we group mass parameters with the spin vectors 
$a_i=\cS_i/m_i$ and consider the expressions to be functions of the ring radii $a_i$.
In this way the generating amplitudes simultaneously make manifest the self-force 
expansion 
as well as the PM expansion \cite{Driesse:2024feo}.  
An example diagram
contributing to $\widehat {\cal M}_s^n(1,2,\ldots, n)$ is displayed in
\cref{fig:waveformEx}.

%--------------------------------------------------------------------------------
\paragraph{Symmetry:} 
%--------------------------------------------------------------------------------
The amplitude $\widehat{\cal M}_s^{n}(1,2,\ldots, n)$ is symmetric under the
exchange of sources,
\begin{align}
\widehat{\cal M}_s^{n}\big[\sigma_n(1),\sigma_n(2),\ldots, \sigma_n(n)\big] = \widehat{\cal M}_s^{n}(1,2,3\ldots, n)\,, 
\end{align}
where $\sigma_n$ denotes a permutation of $n$ objects. This symmetry allows to
reduce the number of diagrams which need to be computed. In particular, one can
symmetrize over individual diagrams to obtain an amplitude.

%--------------------------------------------------------------------------------
\paragraph{Generating amplitudes:}  
%--------------------------------------------------------------------------------
The very same $n$-worldline diagrams may as well be used to obtain higher-order
(classical) gravitational corrections. In fact, by inspection one finds that
identifying worldlines to originate from the same action $S_i$ (with parameters
$m_i$, $b_i$ and $v_i$), leading-order diagrams can be transcribed to
higher-order contributions~\cite{Shen:2018ebu,Mogull:2020sak}.  Consequently, when 
computing an $n$-body waveform, we obtain as well the $\mathcal{O}(G^m)$ corrections
to the  $(n-m)$-body waveform. To give an example, the amplitude
\begin{align}
\widehat{\cal M}_s^{2}(1,2) \sim G^{2} m_1 m_2
\end{align}
contributes at leading order to the two-body interaction, while the amplitudes
\begin{align}
\widehat{\cal M}_s^{3}(1,2,2) &\sim G^{3} m_1 m_2^2  
\qquad \mbox{and} \qquad
\widehat{\cal M}_s^{3}(1,1,2) \sim G^{3}\, m_1^2 m_2
\end{align}
contribute at the next-to-leading order $G^3$. By iteration, all higher-order
corrections arise from the generating amplitudes \eqref{eqn:genAmpl}, 
which justifies this notion.

\begin{figure}[ht!]
 \centering
 \begin{subfigure}[t]{0.32\textwidth}
\centering
  \includegraphics[width=0.8\textwidth]{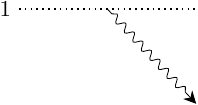}
  \subcaption{}
 \end{subfigure}
 \begin{subfigure}[t]{0.32\textwidth}
  \centering
  \includegraphics[width=0.8\textwidth]{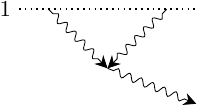}
  \subcaption{}
 \end{subfigure}
 \begin{subfigure}[t]{0.32\textwidth}
  \centering
  \includegraphics[width=0.8\textwidth]{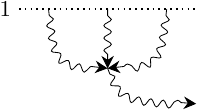}
  \subcaption{}
 \end{subfigure}
 \caption{Example diagrams corresponding to corrections of the static
 background at different orders in $\kappa$. Dotted lines represent background
 fields $\Psi_i^a$, and wavy lines represent graviton fields $h_{\mu\nu}$.}
 \label{fig:staticBG}
\end{figure}

%--------------------------------------------------------------------------------
\paragraph{Static background:} 
%--------------------------------------------------------------------------------
Identifying all background data with the ones of a single source,
\begin{align}
\widehat{\cal M}_s^{n}(1,1,...,1) \sim G^{n}\, m_1^n 
\end{align}
gives static contributions to the metric, which are the leading term in the
waveform's self-force expansion.  Representative diagrams obtained from
this identification are shown in \cref{fig:staticBG}. 
Such diagrams give the perturbative expansion of a black-hole metric translated
by the impact parameter $b_1$ (see e.g. \cite{Cheung:2020gbf,Brandhuber:2021eyq, Chava_2023}), and we will
discuss them further below.

%--------------------------------------------------------------------------------
\paragraph{Sum of contributions:}
%--------------------------------------------------------------------------------
Putting the generating amplitudes back together to obtain the impact-parameter
space amplitude, we have to sum up generating amplitudes of various
multiplicities. For concreteness, we state this for two-body scattering up to
order $G^3$,
\begin{align}\label{eqn:fullAmpl}
\widehat {\cal M}_s &= 
\widehat {\cal M}_s^1(1) +  
\widehat {\cal M}_s^1(2) + 
\widehat {\cal M}_s^2(1,1) +  
\widehat {\cal M}_s^2(1,2) +  
\widehat {\cal M}_s^2(2,2)  \nonumber \\
&+ \widehat {\cal M}_s^3(1,1,1)  +
 \widehat {\cal M}_s^3(1,1,2)  +
 \widehat {\cal M}_s^3(1,2,2)  +
 \widehat {\cal M}_s^3(2,2,2)  + {\cal O}(G^4)\,.
\end{align}
It is important to note that only connected diagrams contribute, since
disconnected diagrams are excluded in the classical waveform observable by
definition in the worldline QFT.  In a field-theory approach connected
worldline diagrams may appear as disconnected field-theory
diagrams~\cite{Bini:2024rsy}. Thus the notion of diagram topology differs in
the two approaches. An example for this are the static background contributions
mentioned above.

%--------------------------------------------------------------------------------
\paragraph{Feynman integrals:}
%--------------------------------------------------------------------------------
The general form of a WQFT diagram closely resembles the one of Feynman
amplitudes.  In particular, a single diagram gives rise to a momentum integral
in dimensional regularization, with a rational integration kernel and
delta-function constraints.  A given diagram of the generating amplitude
$\widehat{\cal M}_s^n$ takes the form
\begin{align}
 D_{n,\,\gamma} = \kappa^{2n-1} \left(\prod_{i=1}^n m_i\right)\, \int d\mu_n(1,\ldots, n) 
 \frac{N_\gamma}{\prod_{i\in P(\gamma)} \rho_i}\;,
  \label{eq:GeneralNDiagram}
\end{align}
where $\rho_i$ denotes retarded propagators labeled by $i$ in the set
$P(\gamma)$ of all edges of the diagram $\gamma$ and $N_\gamma$ is the
polynomial numerator emerging from the Feynman rules. 
The measure factor arises
from energy conservation along the worldlines and momentum conservation in the
graviton interactions. It is given by,
\begin{align}\label{eqn:measure}
 d\mu_n(1,\ldots, n) \equiv \left[ \prod_{j=1}^n \frac{d^dq_j}{(2\pi)^d}\, e^{i q_j\cdot b_j}\, \hat\delta(v_j\cdot q_j) \right] 
 \hat\delta^{(d)}\left(k-\sum_{j=1}^n q_j\right)\;,
\end{align}
where we introduced the shorthand notation $\hat\delta(x)\equiv 2\pi\delta(x)$.
The $q_i$'s refer to the sum of all graviton momenta that are connected to
background $i$.  A set of example diagrams is shown e.g. in \cref{fig:3BodyDiagrams}. 
Each diagram has a specific overall mass scaling and dependence on the 
gravitational coupling $\kappa$. 

The main distinction of a WQFT integral \eqref{eq:GeneralNDiagram} from a Feynman 
integral is the Fourier phase factor depending
on the impact parameters $b_i$.  It is to convenient to split the integration into a Fourier
integration and a remaining standard Feynman amplitude $\cM_s^{n,g}$. For the generating amplitudes
\eqref{eqn:genAmpl}, which include $n$ distinct sources, the Feynman amplitude
is obtained by dropping the measure altogether. For contributions with
identified sources, as we discuss below,  the integration includes Fourier
integrals as well as momentum integrals over rational integration kernels. 
Separating the Fourier integration is technically convenient, since otherwise an 
extension of the standard Feynman integral calculus is required \cite{Brunello:2024ibk}.  
Conceptually, this split makes sense, since the momentum-space amplitude matches the 
ones obtainable in the field-theory approach (see \cref{sec:QFT}).

We now introduce the momentum-space amplitudes.  We start with the generating
amplitudes and define the  momentum-space generating amplitude ${\cal
M}_s^{n,g}(1,2,\cdots, n)$ implicitly through
\begin{align} \label{eqn:genAmplMom}
\widehat{\cal M}_s^{n}(1,\ldots, n)  = 
\int  d\mu_n(1,\ldots, n)  \,
{\cal M}_s^{n,g}(1,\ldots, n)\,,
\end{align}
assuming distinct sources and thus distinct integer labels.

The treatment of integration differs for amplitudes with identified sources.
In this case, not all impact parameters $b_i$ are distinct and we can perform a
linear change of integration variables $q_i$ to isolate integrals that do not
affect the Fourier phases at all,
\begin{align}\label{eqn:measureFactorized}
 d\mu_n(i_1,...,i_{n-m},1,2,\ldots, m) \rightarrow    
 d\mu_m(1,2,\cdots m) \times  
\left[ \prod_{j=1}^{n-m} \frac{d^d\ell_j}{(2\pi)^d}\, \hat\delta(v_{i_j}\cdot \ell_j) \right] \,,
\end{align}
where $i_k \in \{1,\ldots m\}$.  The phase-independent integration directions
are denoted by $\ell_i$ and they match loop integration in the field-theory
approach.  We define momentum-space amplitudes by
\begin{multline} \label{eqn:defWQFTMomAmpl}
{\cal M}_s^{n}(i_1,\ldots,i_{n-m},1,2,\ldots, m)  \\ \equiv
\left[ \prod_{j=1}^{n-m} \int\frac{d^d\ell_j}{(2\pi)^d}\, \hat\delta(v_{i_j}\cdot \ell_j) \right] \,{\cal M}_s^{n,g}(i_1,\ldots,i_{n-m},1,2,\ldots, m)\,,
\end{multline}
such that
\begin{align}\label{eqn:defWQFTMomAmpl2}
\widehat{\cal M}_s^{n}(i_1,\ldots,i_{n-m},1,2,\ldots, m)  = 
\int  d\mu_m(1,\ldots, m)~{\cal M}_s^{n}(i_1,\ldots,i_{n-m},1,2,\ldots, m)\,.
\end{align}
These momentum-space amplitudes are equivalent to  field-theory loop amplitudes with some
propagators being delta functions. In the argument, an ordered set of
indices is sorted to the end and repeated indices $\{i_1,\ldots i_{n-m}\}$ are inserted in the
beginning of the argument list in \cref{eqn:measureFactorized,eqn:defWQFTMomAmpl,eqn:defWQFTMomAmpl2}. 
When unambiguous, we will also treat the labels as symmetric arguments. 

In the case of computing higher-order corrections to the two-body system, we
will see that the resulting integrals are Feynman integrals.  In the remainder of this section,
we analyze which diagrams are required to extract the waveform at order
$\mathcal{O}(G^3)$ and what integrals they will give rise to.

To give an example for these formal steps we consider the static background
contributions.  For these we identify all sources, say with the worldline one,
\begin{align}
\{m_i,v_i,b_i,S_i\} \rightarrow \{m_1,v_1,b_1,S_1\}\;, \quad
q_{i<n} \rightarrow \ell_i 
\quad\mbox{and}\quad 
q_{n} \rightarrow q_1-\sum_{i=1}^{n-1} \ell_i \;,
\end{align}
in \cref{eqn:measure}. The measure factorizes as,
\begin{align}
d\mu_n(1,\ldots, n)&=\int d\mu_1(1) \times \left[\prod_{i=1}^{n-1} \frac{d^d\ell_i}{(2\pi)^d}\, 
\hat\delta(\ell_i\cdot v_1) \right]
\end{align}
and we obtain the momentum-space amplitude,
\begin{align}
{\cal M}_s^{n}(1,1,\ldots ,1)  = 
\left[ \prod_{j=1}^{n-1} \int \frac{d^d\ell_j}{(2\pi)^d}\, \hat\delta(v_{1}\cdot \ell_j) \right] \,{\cal M}_s^{n,g}(1,\ldots,1)\,,
\end{align}
which is an $(n-1)$-loop contribution to the static background contributing at
${\cal O}(G^n)$.  The impact-parameter amplitude is the Fourier transform,
\begin{align}
\widehat{\cal M}_s^{n}(1,1,\ldots ,1)  &= 
\int  d\mu_1(1)  \,{\cal M}_s^{n}(1,1,\ldots,1)\,,\\
d\mu_1(1) &=  
\frac{d^dq_1}{(2\pi)^d}\, e^{i q_1\cdot b_1}\, \hat\delta(v_1\cdot q_1)
 \hat\delta^{(d)}\left(k-q_1\right)\,.
\end{align}
As we have hinted with the choice of variables names $\ell_i$, $n-1$ of the
integrals turn into Feynman-like integrals (albeit with retarded propagators).
Representative diagrams obtained from this identification are shown in
fig.~\ref{fig:staticBG}. The remaining integration over $q_1$ corresponds to
the Fourier transformation from momentum to impact parameter space. The
delta functions enforce the zero-energy condition $v_1\cdot k=0$ and the
diagrams yield only static metric fluctuations that are not detectable by
gravitational-wave
observatories. 
%--------------------------------------------------------------------------------
\subsection{Two-body scattering from seven-point trees} 
\label{sec:IdentificationTheory}
\begin{figure}[ht!]
 \centering
 \begin{subfigure}[t]{0.23\textwidth}
   \centering
   \includegraphics[width=\textwidth]{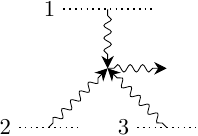}
   \subcaption{}
   \label{fig:diagA}
 \end{subfigure}
 \hfill
 \begin{subfigure}[t]{0.23\textwidth}
   \centering
   \includegraphics[width=\textwidth]{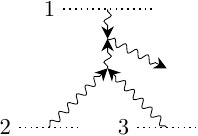}
   \subcaption{}
   \label{fig:diagB}
 \end{subfigure}
 \hfill
 \begin{subfigure}[t]{0.23\textwidth}
   \centering
   \includegraphics[width=\textwidth]{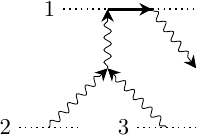}
   \subcaption{}
   \label{fig:diagC}
 \end{subfigure}
 \hfill
 \begin{subfigure}[t]{0.23\textwidth}
   \centering
   \includegraphics[width=\textwidth]{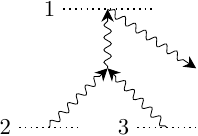}
   \subcaption{}
   \label{fig:diagD}
 \end{subfigure}
 \\
 \begin{subfigure}[t]{0.23\textwidth}
   \centering
   \includegraphics[width=\textwidth]{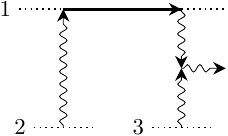}
   \subcaption{}
   \label{fig:diagE}
 \end{subfigure}
 \hfill
 \begin{subfigure}[t]{0.23\textwidth}
   \centering
   \includegraphics[width=\textwidth]{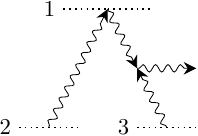}
   \subcaption{}
   \label{fig:diagF}
 \end{subfigure}
 \hfill
 \begin{subfigure}[t]{0.23\textwidth}
   \centering
   \includegraphics[width=\textwidth]{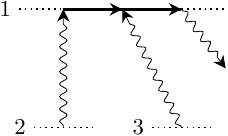}
   \subcaption{}
   \label{fig:diagG}
 \end{subfigure}
 \hfill
 \begin{subfigure}[t]{0.23\textwidth}
   \centering
   \includegraphics[width=\textwidth]{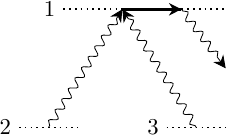}
   \subcaption{}
   \label{fig:diagH}
 \end{subfigure}
 \\
 \begin{subfigure}[t]{0.23\textwidth}
   \centering
   \includegraphics[width=\textwidth]{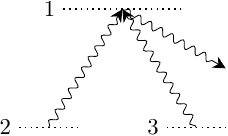}
   \subcaption{}
   \label{fig:diagI}
 \end{subfigure}
 \hfill
 \begin{subfigure}[t]{0.23\textwidth}
   \centering
   \includegraphics[width=\textwidth]{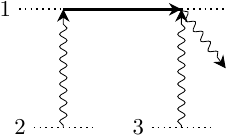}
   \subcaption{}
   \label{fig:diagJ}
 \end{subfigure}
 \hfill
 \begin{subfigure}[t]{0.23\textwidth}
   \centering
   \includegraphics[width=\textwidth]{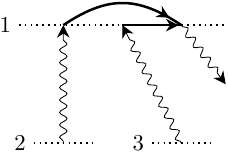}
   \subcaption{}
   \label{fig:forPartialFractioning}
 \end{subfigure}
 \hfill
 \begin{subfigure}[t]{0.23\textwidth}
   \centering
   \includegraphics[width=\textwidth]{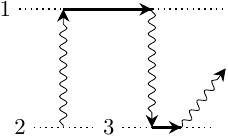}
   \subcaption{}
   \label{fig:diagL}
 \end{subfigure}
 \\
 \begin{subfigure}[t]{0.23\textwidth}
   \centering
   \includegraphics[width=\textwidth]{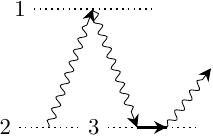}
   \subcaption{}
   \label{fig:diagM}
 \end{subfigure}
 \hfill
 \begin{subfigure}[t]{0.23\textwidth}
   \centering
   \includegraphics[width=\textwidth]{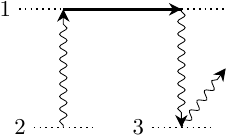}
   \subcaption{}
   \label{fig:diagN}
 \end{subfigure}
 \hfill
 \begin{subfigure}[t]{0.23\textwidth}
   \centering
   \includegraphics[width=\textwidth]{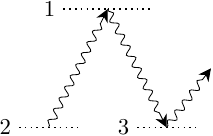}
   \subcaption{}
   \label{fig:diagO}
 \end{subfigure}
 \caption{The diagrams contributing to the leading-order three-body waveform in
the WQFT. Dotted lines represent background fields $\Psi_i^a$, wavy lines
represent graviton fields $h_{\mu\nu}$ and bold black lines depict $z^\mu_i$,
$\phi_i^a$ and $\bar{\phi}_i^a$ fields. Arrows indicate the direction of
momentum flow, relevant for the retarded propagators.
The set of all diagrams is obtained by permuting the labels 
$(1,2,3) \rightarrow (\sigma(1),\sigma(2),\sigma(3))$.
}
 \label{fig:3BodyDiagrams}
\end{figure}

%--------------------------------------------------------------------------------
We now specialize the earlier methods to obtain the next-to-leading order 
corrections,
\begin{align}
{\cal M}_s^{3}(1,1,2) \quad \mbox{and} \quad  {\cal M}_s^{3}(1,2,2) \,,
\end{align}
which contribute at ${\cal O}(G^3)$.  We start by considering the generating
amplitudes ${\cal M}_s^{3,g}(1,2,3)$ of the leading-order three-body
black-hole scattering.  The relevant diagrams at $\mathcal{O}(\kappa^5)$ are
shown in \cref{fig:3BodyDiagrams}. We want to emphasize that the bold lines
correspond to either $z_i^\mu$, $\phi_i^a$ or $\bar{\phi}_i^a$ fields and all
permutations of sources for each diagram have to be taken into account. We also
do not consider disconnected diagrams, since they do not contribute to the
classical waveform observable.

\begin{figure}[ht!]
\centering
 \begin{subfigure}[t]{0.40\textwidth}
   \centering
   \includegraphics[height=2.5cm]{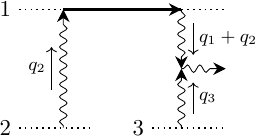}
   \subcaption{}
 \end{subfigure}
\begin{minipage}{.1\textwidth}
\centering
\vspace{-2cm}$\longrightarrow\quad$
\end{minipage}
 \begin{subfigure}[t]{0.40\textwidth}
   \centering
   \includegraphics[height=2.9cm]{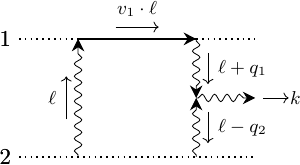}
   \subcaption{}
 \end{subfigure}
 \caption{
Diagram (a) with three distinct sources is relabeled ($3\rightarrow 2$) to obtain
diagram (b) involving two worldlines. While diagram (a) is a LO 
contribution to the gravitational exchange of three sources, diagram (b) contributes at 
NLO to the waveform of two scattering objects.}
 \label{fig:3to2BodyDiagram}
\end{figure}

The NLO correction to the binary scattering is recovered by
identifying two of the background fields to be identical, as shown in
\cref{fig:3to2BodyDiagram}.  The amplitudes' mass dependence ${\cal
M}_s^{3}(1,1,2) \sim  G^3  m_1^2 m_2$ and  ${\cal M}_s^{3}(1,2,2) \sim G^3 m_1 m_2^2$ 
solely arises from the type of identification.
Let us now focus on the $(m_1 m_2^2)$ part ${\cal M}_s^{3}(1,2,2)$ of the
classical amplitude. (${\cal M}_s^{3}(1,1,2)$ follows from relabeling.) Each
diagram contributes multiple terms to the $(m_1m_2^2)$ structure as all possible
permutations of the identification of the background fields have to be performed
for each of the diagrams. In practice, we obtain all permutations of the indices
$(1,2,3)$ for all diagrams shown in \cref{fig:3BodyDiagrams} and then set
\begin{align}
  \{m_3,v_3,b_3,\cS_3\} \rightarrow \{m_2,v_2,b_2,\cS_2\}\;.
  \label{eq:identificationMappingGeneric}
\end{align}

Next, we deal with the momentum integrals.  In order to separate the Fourier
integral we must find a parameterization of $q_2$ and $q_3$, such that the total
momentum emitted from the two worldline-$2$ contributions adds up to $q_2$.  This means the
parameterization must fulfill $q_2+q_3 \rightarrow q_2$.  A family of such transformations is
\begin{align}
  q_2 \rightarrow \pm\ell + x q_2\;,\qquad q_3 \rightarrow (1-x)q_2  \mp \ell\;. \label{eq:mappingGeneric}
\end{align}
Here $x$ is a number that can, together with the sign of $\ell$, be freely
chosen. We will make use of this freedom to simplify the expressions of the
emerging integrals.  In particular, the choice is guided by the goal to obtain
the following integration measure: 
\begin{align}
 \int d \mu_3(1,2,3) \rightarrow\int d \mu_2(1,2) \, \int \frac{d^d \ell}{(2 \pi)^d}\hat{\delta}(\ell\cdot v_2)\;.
 \label{eq:measureM3}
\end{align}
A suitable choice of $x$ and the sign of $\ell$ depends on the specific
diagrams under consideration and will be discussed in
\cref{sec:integrandcomputation}.  The identification procedure maps the tree-level
diagram into a Feynman integral over a loop-momentum $\ell$,
\begin{align}
{\cal M}_s^{3}(1,2,2)  = \int 
\frac{d^d\ell}{(2\pi)^d}\, \hat\delta(v_2\cdot \ell) \,{\cal M}_s^{3,g}(1,2,2)\,.
\end{align}
%
%The Feynman rules give ${\cal M}_s^{n,g}$ in terms of tree diagrams.  
We now
discuss the Feynman integral topologies which need to be computed.

\begin{figure}[ht]
  \centering
  \includegraphics[width=0.3\linewidth]{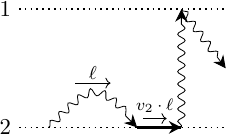}
  \caption{A self-energy diagram that does not contribute as it gives rise to
scaleless integrals.}
  \label{fig:SelfEnergy}
\end{figure}

%--------------------------------------------------------------------------------
\paragraph{Self-energy diagrams:} 
%--------------------------------------------------------------------------------
Some identifications lead to self-energy contributions which we are allowed to omit.
An example diagram is shown in
\cref{fig:SelfEnergy}, which originates from the $(1,2,3)\rightarrow (3,2,1)$
permutation of diagram \cref{fig:diagN} after identifying worldline 3 with worldline 2.
Such diagrams naively appear to be divergent, since they contain a
delta function $\delta(v_2\cdot \ell)$ as well as a propagator
$i/(v_2\cdot\ell+ i \epsilon)$. After representing the delta functions as a
difference of propagators 
\begin{align}\label{eqn:reverseUnitarity}
2\pi i \, \delta(\omegal) = \frac{1}{\omegal - i\epsilon} - \frac{1}{\omegal + i\epsilon} \,,
\end{align}
we obtain scaleless integrals depending on a single propagator raised to higher
power ${1/(v_2\cdot\ell+ i \epsilon)^n}$. Such integrals integrate to zero in
dimensional regularization and we drop them in our calculation.

\begin{figure}[ht!]
 \centering
 \begin{subfigure}[t]{0.48\textwidth}
   \centering
   \includegraphics[height=3cm]{figs/parametrization2.pdf}
   \caption{}
   \label{fig:ParentTopology}
 \end{subfigure}
 \begin{subfigure}[t]{0.48\textwidth}
   \centering
   \includegraphics[height=3cm]{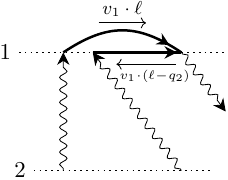}
   \caption{}
   \label{fig:PartialFractioning}
 \end{subfigure}
 \caption{a) The parent topology of the $m_1 m_2^2$ sector of integrals. b)
This topology has linearly dependent propagators for the $z/\phi$-lines and can
be reduced with the help of partial fractioning.}
\end{figure}

%--------------------------------------------------------------------------------
\paragraph{Integral topologies:}
%--------------------------------------------------------------------------------
Upon identification of background worldlines ($3 \rightarrow 2$) in the
permutations $\{\sigma^i_3\}_{i=1,6}$ of $(1,2,3) \rightarrow (\sigma^i_3(1),\sigma^i_3(2),\sigma^i_3(3))$
of \cref{fig:3BodyDiagrams}, we encounter two integral families, defined by the
pentagon diagram \cref{fig:ParentTopology} and its pinches, as we will now
discuss.
The general form of these pentagon integrals is
\begin{equation}
\mathcal{I}_\nuv = \int \frac{d^d \ell}{(2\pi)^d} 
\frac{1}{\rho_1^{\nu_1} \rho_2^{\nu_2} \rho_3^{\nu_3} \rho_4^{\nu_4}} \delta(\rho_5)\;,
\label{eq:genericIntegrand} 
\end{equation}
with $\nuv=\{\nu_1,\nu_2,\nu_3,\nu_4\}$ and
\begin{align}
  &\rho_1 = \ell^2 + i \epsilon \,\mathrm{sgn}(\ell^0) \;, \quad \rho_2 = v_1\cdot \ell + i \epsilon\;, \quad
   \rho_3 = (\ell+q_1)^2+i \epsilon\, \mathrm{sgn}(\ell^0+q_1^0)\; \nn\\
  &\rho_4 = (\ell-q_2)^2-i\epsilon\,\mathrm{sgn}(\ell^0-q_2^0)\;,\quad \rho_5 = v_2\cdot\ell \;.
 \label{eq:IntegrandPropagators}
\end{align}
While not originating directly from a propagator in a Feynman diagram, we
include the $\ell$-dependent delta function $\delta(\rho_5)$ from momentum
conservation.  Originally, it is associated to the measure \eqref{eqn:measure}, 
but is regrouped after identification of the background lines
\eqref{eqn:measureFactorized}. 

Most diagrams are obviously identified as pinches of the above pentagon.  The
only  non-trivial case arises from identifications of
\cref{fig:forPartialFractioning} and the topology is shown in
\cref{fig:PartialFractioning}. The corresponding topology is given by
\begin{equation}
\mathcal{I}' = \int \frac{d^d \ell}{(2\pi)^d} 
\frac{1}{\rho_1\rho_2 \rho_6 \rho_4} \delta(\rho_5)\;, \quad\mbox{with}\quad  
\rho_6 = v_1\cdot (\ell - q_2) - i \epsilon\;.
 \label{eq:IntegrandPropagatorsDep}
\end{equation}
In this topology, the propagators $\rho_2$ and $\rho_6$ associated with the
$z_i^\mu$ or $\phi_i^a$ fields are linearly dependent. Their product can be simplified by
partial fractioning,
\begin{align}
\frac{1}{[v_1\cdot \ell + i \epsilon ] [ v_1\cdot (\ell-q_2) -  i \epsilon )]} = 
-\frac{1}{v_1\!\cdot\! q_2}\left[
\frac{1}{[v_1\cdot \ell + i \epsilon ]}
-\frac{1}{[v_1\cdot (\ell-q_2) - i \epsilon ]}
\right]\;.
\label{eqn:partialFraction}
\end{align}
Inserting these expression we can express the pentagon integral $\cI'$ in terms of two
box integrals of the pentagon family $\mathcal{I}$,
\begin{align}
\mathcal{I}' =  -\frac{1}{v_1\!\cdot\! q_2} 
	\left[ 
	\mathcal{I}_{1,1,0,1} + 
	\mathcal{I}_{1,1,0,1}\Big|_{\ell\rightarrow q_2-\ell }
	\right]\,.
\end{align}
Similarly, pinches of $\mathcal{I}'$ are mapped to the pentagon family of
$\mathcal{I}$.  Using integration-by-parts (IBP) relations
\cite{Tkachov:1981wb,Chetyrkin:1981qh,Laporta:2000dsw} we can reduce all tensor
integrals to scalar master integrals in the pentagon family.  The second
integral family (appearing in contributions of mass scaling $m_1^2 m_2$) is
recovered from exchanging $(1\leftrightarrow2)$.

In summary, we require the integrands of ${\cal M}^{3,g}_s(1,2,3)$ as well as
the integrals $\mathcal{I}$ for the momentum-space amplitudes 
${\cal M}^{3}_s(1,2,2)$ and ${\cal M}^{3}_s(1,1,2)$. We will turn to their
computation next.

%--------------------------------------------------------------------------------
\section{Waveform computation}
\label{sec:computation}
%--------------------------------------------------------------------------------
In this section, we will describe the technical details of our computation of
the spectral waveform. We will start by discussing the construction of the
LO three-body integrand and explain the symmetrization and
identification of the background fields. Afterwards, we elaborate on how we
map this into higher-order corrections for the two-body waveform. Finally, we
will discuss the evaluation of the integrals with retarded propagators.

%-------------------------------------------------------------------------------
\subsection{Amplitude computation}
\label{sec:integrandcomputation}
%-------------------------------------------------------------------------------
We start by calculating the LO three-body momentum-space amplitude
$\mathcal{M}^{3,g}_s(1,2,3)$.  
The diagrams that contribute are shown in \cref{fig:3BodyDiagrams}.
We generate them with \texttt{qgraf} \cite{Nogueira:1991ex},
including all their symmetrizations and symmetry factors.  The vertex Feynman rules
were computed from \cref{eq:FullAction} with the help of
\texttt{xAct} \cite{xAct, Nutma:2013zea}. We import the
\texttt{qgraf}-diagrams, plug in Feynman rules and perform summation over
indices with the help of \texttt{alibrary} \cite{Magerya:2025}. Finally, we
replace all scalar products with the minimal set of variables introduced in
\cref{sec:notation}.

In the next step, we generate the two-body NLO integrand 
from the generating three-body amplitude $\mathcal{M}^{3,g}_s(1,2,3)$.
For that, we must set the background data of two black holes to the same initial condition, 
thus identifying them with each other. In this way we obtain
$\mathcal{M}^{3}_s(1,2,2)$ and $\mathcal{M}^{3}_s(1,1,2)$.

Next, we want to map the momenta in such a way, that the emerging integrals are
mapped to the pentagon integral family \eqref{eq:genericIntegrand} shown in
\cref{fig:ParentTopology}. 
For simplicity we focus on the amplitude $\mathcal{M}^{3}_s(1,2,2)$ with mass
scaling $m_1 m_2^2$.
At a diagrammatic level we identify
\begin{equation}
 \{m_3,v_3,b_3,\cS_3\} \rightarrow \{m_2,v_2,b_2,\cS_2\}\;,
\end{equation}
to find all diagrams and thus the contributing Feynman amplitudes. 

\begin{figure}[ht!]
	\centering
	\begin{subfigure}[t]{0.32\textwidth}
		\centering
		\includegraphics[width=0.8\textwidth]{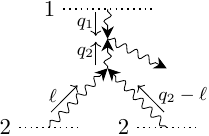}
		\subcaption{}
		\label{fig:mapping1}
	\end{subfigure}
	\begin{subfigure}[t]{0.32\textwidth}
		\centering
		\includegraphics[width=0.8\textwidth]{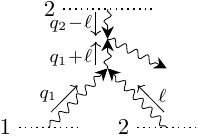}
		\subcaption{}
		\label{fig:mapping2}
	\end{subfigure}
	\caption{Example diagrams after momentum maps $L_1$ and $L_2$ were 
	applied to relate their momenta to the ones of the pentagon family.}
	\label{fig:mappingExamples}
\end{figure}

In order to map all integrals to 
the pentagon family  
\eqref{eq:genericIntegrand}
we use the linear map \eqref{eq:mappingGeneric} for $q_2$ and $q_3$.
It turns out that two particular choices suffice to bring all diagrams 
of \cref{fig:3BodyDiagrams} (and their symmetrizations) to the desired form:
\begin{align}
	L_1:\quad&  q_2 \rightarrow \ell\;, \quad q_3 \rightarrow q_2 - \ell \,;\\
	L_2:\quad&   q_2 \rightarrow q_2-\ell\;, \quad q_3 \rightarrow \ell\,.
\end{align}
To make this clear, we next discuss an example diagram for each of the two
mappings.  Map $L_1$ is suitable for diagram \ref{fig:diagB} after the relabeling
$(1,2,3)\rightarrow (1,2,2)$ of worldline data. The relabeled diagrams are shown
in \cref{fig:mapping1}.  For an application of map $L_2$, we also consider
diagram \ref{fig:diagB}, but for the label permutation $(1,2,3)\rightarrow
(2,1,2)$. By comparing the propagator momenta of \cref{fig:mapping2}, it is
clear that they match the ones of the pentagon family
\eqref{eq:genericIntegrand}.
%It is easy to figure out which mapping to use for each diagram.
The remaining diagrams follow in a similar way.

Technically, it is convenient to perform the momentum maps at the level of invariants.
%However, we  it is more convenient to replace directly at the level of scalar
%products.
Considering \cref{eq:3BodyVariables} we find that $9$ out of the $14$ variables
map directly to the $5$ variables of \cref{eq:2BodyVariables}. Those
are
\begin{gather} \omega_1\;, \qquad \omega_2\;, \qquad \omega_{31} \to
\omega_2\;, \qquad \omega_3 \to \omega_2\;,\qquad q_1^2\;, \qquad q_{23}^2 \to q_2^2\;, \nonumber \\ y_{12}
\to y\;, \qquad y_{23}\to 1\;, \qquad y_{13} \to y\;.  \end{gather}
The remaining variables of the three-body scattering process
can be written in terms of the external invariants or propagators of the two-body amplitude. This map
then depends on the specific maps ($L_1$ or $L_2$).
The respective replacement rules are
\begin{align}
L_1:\quad & \omega_{12} \rightarrow \rho_2\;, \quad
\omega_{23} \rightarrow -\rho_5\;, 
\quad q_2^2\rightarrow \rho_1\;,
\quad q_3^2\rightarrow \rho_4\;, \nonumber \\
& q_{12}^2 \rightarrow \rho_3\,; \\
L_2:\quad & \omega_{12}\rightarrow \omega_1 - \rho_1\;,\quad 
\omega_{23} \rightarrow \rho_5 \;, 
\quad q_2^2 \rightarrow \rho_4\;,
\quad q_3^3 \rightarrow \rho_1\;, \nonumber \\
& q_{12}^2 \rightarrow q_1^2 - q_2^2 + \rho_1 - \rho_3 + \rho_4\;. 
\end{align}
Performing these replacements on the diagrams of the three-body amplitude yields the 
integrand of the two-body momentum-space amplitude 
$\mathcal{M}^{3}_s(1,2,2)$ in terms of invariants and propagators. 
To simplify the expressions after the mapping to these loop integrands, we exploit 
diagram isomorphisms under transformations of $\ell$ and collect matching diagrams.
Finally, the resulting tensor integrals were reduced to scalar integrals with
the help of IBP relations using \textsc{LiteRed}~\cite{Lee:2012cn,Lee:2013mka}.
The amplitude $\mathcal{M}^{3}_s(1,1,2)$ is obtained by relabeling.

%-------------------------------------------------------------------------------
\subsection{Integrals}
\label{sec:integrals}
%--------------------------------------------------------------------------------
As a next step in the waveform computation, we must calculate the integrals in
\cref{eq:genericIntegrand}. In particular, we have to consider all scalar master integrals
$\cI_\nuv$ with indices $\nu_i=0,1$. 
As above we focus on the integral family of $M_s^{3}(1,2,2)$ and 
obtain the integrals of $M_s^{3}(1,1,2)$ by relabeling.

So far, we have neglected to discuss the $i\epsilon$ prescription of the
propagators. This aspect is the main focus of this section. For each propagator
of the diagrams in \cref{fig:3BodyDiagrams}, the arrow shows a flow of
causality towards the outgoing gravitational wave, which in turn specifies the retarded 
propagator prescriptions \eqref{eq:zProp} and \eqref{eq:phiProp}. As we discuss next, 
we find that, after integral reduction, these worldline integrals map onto the known 
field‑theory Feynman integrals for the NLO waveform 
\cite{Brandhuber:2023hhy,Herderschee:2023fxh,Caron-Huot:2023vxl,Bohnenblust:2023qmy}.

The building block of the field-theory integrals is the Feynman
integral 
\def\rhof{{d}}
\begin{align}
  \cIh_{\nuv} = \int \frac{d^d\ell}{(2\pi)^d} 
  \frac{1}{\rhof_1^{\nu_1} \rhof_2^{\nu_2} \rhof_3^{\nu_3} \rhof_4^{\nu_4}\rhof_5^{\nu_5}}\;,
\end{align}
where
\begin{align}
	&\rhof_{1} = \ell^2 + i \epsilon \;, \quad\quad \rhof_2 = v_1\cdot \ell + i \epsilon\;, \quad\quad\quad
	\rhof_3 = (\ell+q_1)^2+i \epsilon\,, \; \nn\\
	&\rhof_4 = (\ell-q_2)^2+i\epsilon \;,\quad\quad \rhof_5 = -v_2\cdot\ell+i \epsilon \;.
	\label{eq:IntegrandPropagatorsFeyn}
\end{align}
and we assume $\nu_i=0,1$ below.
Note that these propagators are equivalent to the WQFT ones \eqref{eq:genericIntegrand} up to the i$\epsilon$ prescription.
The integrals $\cIh$ are used to define the waveform master integrals $\cIh^{\sigma_1,\sigma_2}_\nuv$ \cite{Caron-Huot:2023vxl} as the superposition
\begin{align}
 \cIh^{\sigma_1,\sigma_2}_\nuv &\equiv  \cIh_{\nuv}  +(-1)^{\nu_2} \sigma_1 \cIh_{\nuv}  \big|_{v_1\to -v_1}+ (-1)^{\nu_5}\sigma_2 \cIh_{\nuv}  \big|_{v_2\to -v_2} \nn \\
 &+ (-1)^{(\nu_2+\nu_5)}\sigma_1 \sigma_2  \left(\cIh_{\nuv}  -  \mathrm{Cut}_{\rhof_2,\rhof_5}[\cIh_{\nuv} ] \right) \big|_{(v_1,v_2)\to (-v_1,-v_2)}\;,
 \label{eq:IntegralSuperposition}
\end{align}
where the cut-contribution is non-vanishing only for $\nu_2=\nu_5=1$, and defined by
\begin{align}\label{eqn:cut}
\mathrm{Cut}_{\rhof_2,\rhof_5}[\cIh_{\nuv} ] =
\int \frac{d^d\ell}{(2\pi)^d} 
\frac{1}{\rhof_1^{*\nu_1}\rhof_3^{\nu_3} \rhof_4^{\nu_4}}[2\pi i\, \delta(\rhof_2)] [2\pi i\, \delta(\rhof_5)]\,.
\end{align}

We now show how the WQFT integrals $\cI$ of \cref{eq:genericIntegrand}
are equivalent to the integrals $\cIh^{+-}_\nuv$ and $\cIh^{-+}_\nuv$. 
We start with discussing the (in)sensitivity of the
graviton propagators on their $i\epsilon$-prescription and, in a second step, 
we show that the linear-propagator dependence matches.

%-------------------------------------------------------------------------------
\paragraph{Graviton propagators:}
%-------------------------------------------------------------------------------
The graviton propagators of the pentagon integral family
\eqref{eq:genericIntegrand}
 are $\rhof_1$, $\rhof_3$ and $\rhof_4$, as defined in
\cref{eq:IntegrandPropagators}.  For those, the field-theory integrals $\cIh$
use the Feynman prescription \eqref{eq:IntegrandPropagatorsFeyn}, while the WQFT
sets retarded propagators. Nevertheless, for the relevant parent
topology, and hence for all its daughters, those two prescriptions give 
identical results.

To demonstrate this, we carefully study the momentum-conservation conditions 
\begin{equation}
 \delta(v_2\cdot\ell)\,\delta(v_2\cdot q_2)\,\delta(v_1\cdot q_1) \,\delta^{(d)}(k - q_1 - q_2)\;,
\end{equation}
of the diagram shown in \cref{fig:ParentTopology}, as generated by the Feynman
rules. Furthermore we use that the energy of the produced gravitational wave is  positive,
\begin{equation}
 k^0 \ge 0\,.
\end{equation}
First, we recall that the $i\epsilon$-prescription of the retarded graviton
propagators $\{\rho_1,\rho_3,\rho_4\}$ depends on the sign of the energy component flowing
in direction of causality.  
We now make the convenient frame choice, where $v_{2} = (1,0,0,0)$. The
delta functions enforce vanishing energy components 
\begin{align}
v_2\cdot\ell = \ell^0 = 0 \quad \mbox{and} \quad  v_2\cdot (\ell-q_2) = (\ell-q_2)^0=0
\end{align}
for the propagators $\{\rho_1,\rho_4\}$, which are adjacent to worldline 2. 
In the cases
where the energy component is $0$, the $i \epsilon$-dependence of the retarded
propagator drops out completely, but that is not an issue. In such a case, the
only possible divergent phase-space point must have vanishing momentum. 
Such a zero-dimensional singularity is integrable in
$4-2 \epsilon$ dimensions and the integrals do not require this contour deformation. 
Adding any $i
\epsilon$-prescription to the propagator will thus not change the result of the
evaluated integral and we may choose the Feynman-$i\epsilon$ prescription. Next, we turn our
attention to the graviton propagator $\rho_3$. The respective graviton energy
is positive,
\begin{align}
 (\ell+q_1)^0 = (\ell-q_2+k)^0 = k^0 \ge 0\,.
\end{align}
We may thus replace this retarded propagator by the appropriate Feynman propagator
\begin{equation}
 \frac{1}{\rho_3} = \frac{1}{(\ell+q_1)^2 +i \epsilon \, \mathrm{sgn}(\ell^0+q_1^0)} \rightarrow \frac{1}{d_3}=\frac{1}{(\ell+q_1)^2 +i \epsilon} \;.
\end{equation}
Putting these observations together we can use the Feynman-$i\epsilon$
prescriptions for all graviton propagators in the integral family $\cI_\nuv$,
\begin{align}\label{eqn:iepsInvariance}
	\{\rho_1,\,\rho_3,\,\rho_4\} \rightarrow \{\rhof_1,\rhof_3,\rhof_4\}\;.
\end{align}
Finally, we observe that we can use a uniform 
$i\epsilon$-prescription for all graviton propagators of $\cI_\nuv^{\pm\mp}$. 
The only source of a mixed $i\epsilon$-prescription 
is $1/\rhof_1^*$ of the cut contribution \eqref{eqn:cut}.
However, the delta functions of the cut integral \eqref{eqn:cut} imply, like above, that the momentum of 
$\rhof_1$ is space-like. Consequently, the respective propagator pole has vanishing support, and  
the $i\epsilon$-prescription of $1/\rhof_1^*$ is irrelevant. In this way we are allowed to replace
\begin{align}
	\rhof_1^* \quad\rightarrow\quad \rhof_1 \;,
\end{align}
such that we may use Feynman propagators for all gravitons in $\cIh$. 

%-------------------------------------------------------------------------------
\paragraph{Linear propagators:}
%-------------------------------------------------------------------------------
Having determined that the retarded graviton propagators can simply be
replaced by Feynman propagators, we can factor them out of the
sum~\eqref{eq:IntegralSuperposition}.  We next turn our attention to the matter
propagators $\rhof_2$ and $\rhof_5$, and show that they in fact match the
ones of $\cI$. For simplicity, we focus on the indices $\nu_2=\nu_5 = 1$. 
(The reasoning simplifies for lower values of the indices $\nu_{2,5}$.) 
The matter propagators transform under the sign change
of the velocities as
\begin{align}
  \rhof_2  \xrightarrow{v_1 \to - v_1} -\rhof_2^{*}\,,\qquad \rhof_5  \xrightarrow{v_2 \to - v_2} -\rhof_5^{*}\;,
\end{align}
such that the field-theory integrals $\cIh$ turn into
\begin{align}
    \cIh_\nuv^{\sigma_1,\sigma_2} &= \int \frac{d^d \ell}{(2\pi)^d} \frac{1}{\rhof_1^{\nu_1}\rhof_3^{\nu_3} \rhof_4^{\nu_4}  }\left( \frac{1}{\rhof_2 \rhof_5}+ \frac{\sigma_1}{\rhof_2^{*}\rhof_5}+\frac{\sigma_2 }{\rhof_2 \rhof_5^{*}}+\sigma_1 \sigma_2 \Big[\frac{1}{\rhof_2^{*}\rhof_5^{*}}-\left(\frac{1}{\rhof_2^{*}}-\frac{1}{\rhof_2}\right)\left(\frac{1}{\rhof_5^{*}}-\frac{1}{\rhof_5}\right)  \Big] \right)\nn\\
    &= \int \frac{d^d \ell}{(2\pi)^d} \frac{1}{\rho_1^{\nu_1} \rho_3^{\nu_3}\rho_4^{\nu_4}  }\left(\frac{1-\sigma_1 \sigma_2}{\rhof_2 \rhof_5} + \frac{\sigma_1(\sigma_2+1)}{\rhof_2^{*}\rhof_5}+\frac{\sigma_2(\sigma_1+1)}{\rhof_2 \rhof_5^{*}}  \right)\,.
\end{align}
Here we used \cref{eqn:reverseUnitarity} to express delta functions in terms of propagators.
Inserting $\sigma_1=-\sigma_2=1$ we find, 
\begin{align}
    \cIh^{+-}_\nuv = -2 \int \frac{d^d \ell}{(2\pi)^d} \frac{1}{\rhof_1^{\nu_1} \rhof_2 \rhof_3^{\nu_3}\rhof_4^{\nu_4}  } [2\pi i\, \delta(\rhof_5)]\,.
\end{align}
Given that $\rho_2 = \rhof_2$, the invariance \eqref{eqn:iepsInvariance} and the fact that the
delta function $\delta(\rhof_5)$ is insensitive to the
$i\epsilon$-prescription, this integral is proportional to the scalar integral
$\cIh_\nuv$ \eqref{eq:genericIntegrand}.  Integrals
belonging to the family with mass scaling $m_1^2 m_2$ can be found from
exchanging $(1\leftrightarrow2)$ (or equivalently $\sigma_1=-\sigma_2=-1$).

We have thus shown that the WQFT computation requires the same integrals as the waveform
calculation in field theory. A complete list of the required integrals was
provided in ref.~\cite{Caron-Huot:2023vxl} (see also \cite{Bohnenblust:2023qmy}).

%--------------------------------------------------------------------------------
\section{Spinning waveforms from QFT amplitudes}
\label{sec:QFT}
%--------------------------------------------------------------------------------
In WQFT, the classical observables are directly
captured by tree-level matrix elements. Alternatively, we can obtain
the same observables from scattering amplitudes computed in
a field theory of gravity and massive matter by taking a classical limit
\cite{Kosower:2018adc,
Maybee:2019jus, Cristofoli:2021vyo}. 
(This equivalence of WQFT and field-theory is expected on general 
grounds \cite{Damgaard:2023vnx}.) 
In the past, we exploited
the field-theory  approach already for computing the
linear-in-spin corrections to the waveform at NLO in the PM expansion
\cite{Bohnenblust:2023qmy}. 
Here,
we will utilize this approach to cross check the generating WQFT amplitudes
which make up the NLO WQFT integrand. Interpreted differently, we demonstrate a 
direct way to obtain classical integrands from field theory. 
We remark that hyper-classical terms do not appear in this construction. 

For this purpose we employ the following field theory to capture the classical
scattering 
\begin{align}
S= S_{\textrm{EH}} + S_{\rm gf} + \sum_{i=1}^3 S_{(\phi_i,m_i)} + \sum_{i=1}^3 S_{(V_i,m_i)}\;,
\label{eqn:EFT}
\end{align}
with
\begin{align}
 S_{(\phi,m)} &= \frac{1}{2}\int d^4x\, \sqrt{|g|}\left[g^{\mu\nu}(\partial_\mu\phi)(\partial_\mu\phi) - m^2\phi^2\right]\;, \\
 S_{(V,m)} &= -\frac{1}{4}\int d^4x\, \sqrt{|g|}\left[g^{\mu\nu}g^{\alpha\beta} F_{\mu\alpha}F_{\nu\beta} - 2m^2 g^{\mu\nu}V_\mu V_\nu\right]\;, 
\end{align}
where $F_{\mu\nu} = \partial_\mu V_\nu - \partial_\nu V_\mu$.  In this theory
we study the following scattering processes
\begin{equation}
\begin{split}
\textrm{i}: \quad &\phi_1(p_1) + \phi_2(p_2)  +\phi_3(p_3) \rightarrow \phi_1(p_1') + \phi_2(p_2')  +\phi_3(p_3')  + h(k^{s} )\;,\\
\textrm{ii}: \quad &V_1(p_1^{s_1}\, ) + \phi_2(p_2)  +\phi_3(p_3) \rightarrow V_1(p_1'^{s_1'}\, ) + \phi_2(p_2')  +\phi_3(p_3')  + h(k^{s} )\;,\\
\textrm{iii}: \quad &V_1(p_1^{s_1}\, ) + V_2(p_2^{s_2}\, )  +\phi_3(p_3) \rightarrow V_1(p_1'^{s_1'}\, ) + V_2(p_2'^{s_2'}\, )  +\phi_3(p_3')  + h(k^{s} )\;,
\label{eq:CaravelProcess}
\end{split}
\end{equation}
where the superscripts $s, s_i$ and $s_i^\prime$ denote the spin quantum numbers.
These scattering amplitudes allow us to capture up to quadratic-in-spin
contributions~\cite{Vaidya:2014kza} to the scattering of three spinning black
holes. To recover the spin of black hole $3$, we can
permute labels in \cref{eq:CaravelProcess}. Furthermore, we parameterize the
kinematic of these processes via
\begin{align}
 p_1 &= \barm_1 v_1 + \frac{q_1}{2} \;, \quad p_1^\prime = \barm_1 v_1 - \frac{q_1}{2}\;, \quad
 p_2 = \barm_2 v_2 + \frac{q_2}{2} \;, \quad p_2^\prime = \barm_2 v_2 - \frac{q_2}{2}\;, \nn \\
 p_3 &= \barm_3 v_3 + \frac{q_3}{2} \;, \quad p_3^\prime = \barm_3 v_3 - \frac{q_3}{2}\;, \quad k = q_1+q_2+q_3\;.
 \label{eqn:momenta}
\end{align}
The LO scattering amplitude, which we aim to relate to the
generating amplitude $\cM_s^{n,g}$ \eqref{eqn:genAmplMom}, is defined by
\begin{multline}\label{eqn:polAmpl}
\langle {p_1'}^{s_1'},\ldots ,{p_n'}^{s_n'},k^s| T | p_1,\ldots,p_n  \rangle\big|_{\rm tree\,diagrams} = \\
 \quad  (2\pi)^d\delta^d\left[p_1+\ldots +p_n - (k+p_1'+ \ldots +p_n')\right] M^{\rm tree,n}_{\vec s}
\end{multline}
for the $T$-matrix convention $S=1+i T$ and $\vec s=\{s,s_1,s_1'\ldots
,s_n,s_n'\}$. The tree-level amplitude $M^{\rm tree,n}_{\vec s}$ depends
linearly on the polarization states, and in particular $M^{\rm
tree,7}_{\vec s}$ on the choices in \cref{eq:CaravelProcess}.

The most subtle piece of the computation is recovering a classical spin
dependence from the amplitude $M^{\rm tree,n}_{\vec s}$ written in terms of spin-$1$ 
polarization vectors.
We use the Pauli-Lubanski operator (see e.g. \cite{Sexl:1976pg})
given as
\begin{align}
\mathbb{S}_{\mu}=\frac{1}{2m}\varepsilon_{\mu\nu\alpha\beta}\, p^\nu \mathbb{M}^{\alpha\beta}  \;,
\label{eqn:PLOperator}
\end{align}
where $\mathbb{M}^{\rho\sigma}$ are the Lorentz-group generators.  It is equal
to the relativistic spin operator for a state with momentum $p$ and mass $m$
($p^2=m^2$).  Two of the amplitudes $M_{\vec s}^{{\rm tree},7}$ for the processes listed in
\cref{eq:CaravelProcess} will depend on spin$-1$ polarization vectors
$\varepsilon_v^{\mu}(p_i)$. To express those in terms of the Pauli-Lubanski operator
we use a Clebsch-Gordan decomposition of a product of polarization
states~\cite{Cangemi:2022abk},
\begin{align}\label{eqn:polDec} \overline{\varepsilon}_{v'}^{\mu}(p)
\varepsilon_{v}^{\nu}(p) = &\frac{1}{3}
\overline{\varepsilon}_{v'}(p)\cdot\mathbb{P}\cdot\varepsilon_{v}(p)\,\left(\eta^{\mu\nu}-\frac{p^\mu
p^\nu}{m^2} \right)- \frac{i}{2m}\varepsilon^{\mu \nu \rho \sigma}p_{\rho}
\overline{\varepsilon}_{v'}(p)\cdot\mathbb{S}_{\sigma}\cdot\varepsilon_{v}(p) \nonumber \\
& +\overline{\varepsilon}_{v'}(p)\cdot\mathbb{S}^{\{\mu}\mathbb{S}^{\nu\}}
\cdot\varepsilon_{v}(p) \,,
\end{align}
where we defined
\begin{equation}
 \mathbb{S}^{\{\mu}\mathbb{S}^{\nu\}} \equiv 
 \frac{1}{2}\left( \mathbb{S}^\mu \mathbb{S}^\nu + \mathbb{S}^\nu \mathbb{S}^\mu\right)
 -\frac{1}{3}\left(\mathbb{S}\cdot\mathbb{S}\right)\mathbb{P}^{\mu\nu}\;,
\end{equation}
and
\begin{equation}
 \mathbb{P}^{\mu\nu} \equiv \eta^{\mu\nu} - \frac{p^\mu p^\nu}{m^2}\;.
\end{equation}
This decomposition can be obtained for polarization vectors which are defined for
the same momentum. However, the amplitudes $M_{\vec s}^{{\rm tree},7}$ depend on polarization vectors
given by incoming and outgoing momenta. Those polarization states differ
by a Lorentz transformation.  A boost that relates two generic momenta $p$
and $p'$, which both belong to the same mass shell, is given by
\begin{multline}
\Lambda^\mu_{~\nu}(p',p) = \delta^\mu_{~\nu}  \\ + \frac{1}{(p-p')^2-4m^2}
\left[2\frac{(p-p')^2}{m^2}p'^\mu\, p_{\nu} + 4(p-p')^\mu\,p_{\nu} - 2\left(p+p'\right)^\mu (p-p')_\nu\right]\;,
\label{eqn:boost}
\end{multline}
where the two momenta $p$ and $p'$ fulfill $p^2 = p'^2=m^2$. While the scalar
contribution and linear-in-spin correction are insensitive to the frame choice,
we must boost to the "central" frames related to $\bar p_i$ to recover the
correct quadratic-in-spin dependence \cite{Akpinar:2024meg}.  Since $p_i^2 \neq
\bar p_i^2$, we cannot relate the momenta directly by a Lorentz boost and we
define
\begin{align}
    \bar p'_i = \frac{m_i}{\barm_i}\bar p_i\;.
\end{align}
The classical spin dependence of the amplitude is therefore obtained by the replacement
\begin{equation}
\frac{\overline{\varepsilon}_{v}(\bar{p}'_i)\cdot\mathbb{S}^{\mu}\cdot\varepsilon_{v}(\bar{p}'_i)}{\overline{\varepsilon}_{v}(\bar{p}'_i)\cdot\varepsilon_{v}(\bar{p}'_i)} 
\to \cS^{\mu}_i  \quad\mbox{and}\quad  M_{\vec s}^{\rm tree,7} \to M_s^{\rm tree,7}( \cS )  \,,
\end{equation}
after relating the polarization states of the momenta $\bar p'_i$ and $p_i$ by a boost. 

To extract the classical information and remove any quantum corrections, we
take the classical point-mass limit via $\barm_i \rightarrow \infty$, which we implement
through scaling $\barm_i \rightarrow x\, \barm_i'$ and performing a series
expansion around infinite $x$,
\begin{align}
%x \rightarrow \infty \quad \mbox{with} \quad  \left\{\barm_i', y\,,q^2_i\,, q\,,\omega_i\,, \frac{S_i}{x} \,, \sqrt{x}\kappa\right\} = \mbox{fixed} \;.
x \rightarrow \infty \quad \mbox{with} \quad  \left\{\barm_i', y_{ij}\,,q^2_i\,, q_{ij}^2\,,\omega_i\,, \omega_{ij}\,, \frac{\cS_i}{x} \,, \sqrt{x}\kappa\right\} = \mbox{fixed} \;.
\label{eqn:classLimit}
\end{align}
The seven-point tree-level amplitudes $M_s^{{\rm tree},7}$ scale as
\begin{align}
  M_s^{{\rm tree},7}(\cS) \sim \kappa^5 x^5 (\cS/x)^{n_{\cS}} + \mbox{sub-leading terms}  \;,
\end{align}
where $n_{\cS}$ is the order of the spin multipole contribution.  
The classical amplitude is identified as the leading term
in the scaling limit,
\begin{align}
  {\cM}_s^{{\rm tree},7}(\cS) = M_s^{{\rm tree},7}(\cS)\Big|_{\mbox{leading x-scaling}}\,.
\end{align}
The classical field theory amplitude is expected to match the
WQFT momentum-space amplitude
\begin{align}\label{eqn:compAmpls}
\cM_s^{3,g} = \frac{1}{8\, m_1 m_2 m_3} \cM_s^{{\rm tree},7}(\cS),
\end{align}
which we confirm. The relative factor between the 
amplitudes in \cref{eqn:compAmpls}
is compensated by the phase-space measure of the KMOC formalism, such that 
the impact-parameter waveforms match identically.
We would like to emphasise that $M_s^{{\rm tree},n}$ is a field theory tree-level amplitude, 
which allows to construct the waveform integrand. It matches the WQFT generating amplitudes 
and is free of hyper-classical contributions. Furthermore, the construction
generalizes to higher multiplicity and, after source identification, to higher
order corrections.

An extra subtlety in this approach is obtaining the Casimir corrections ($\sim
\cS^2$).  The issue arises because at the level of operators, we may make the
replacement $\mathbb{S}^2 \rightarrow -s(s+1)\mathbb{1}$, which moves terms of
the quadratic-in-spin correction to quantum corrections of the scalar
contribution, and is known as the \textit{Casimir ambiguity}. We employ the
\textit{spin interpolation} method \cite{Akpinar:2024meg}, to extract
the spin Casimir contributions by interpolating between spinless and spinning
amplitudes.
In more detail, working in a $s=1$ theory, we  add a term
$\frac{c}{x^2}(\cS^2 + 2)$ with an undetermined coefficient $c$. We take advantage of
\textit{spin universality} to fix said coefficient; by comparing to a
calculation done in an $s=0$ theory, we require the leading-order and first
quantum correction to be equivalent to the scalar contribution obtained in the
$s=1$ calculation, which uniquely determines the coefficient $c$, and thus also
fixes the Casimir contribution.

\paragraph{Implementation:} For the comparison with the WQFT seven-point 
tree-level amplitudes, a numerical
result of the classical field-theory amplitude is sufficient. We perform the numerical
computation of the field-theory amplitudes 
$M^{\rm tree,n}_{\vec s}$ \eqref{eqn:polAmpl}
within the \Caravel{} framework
\cite{Abreu:2020xvt}. This requires a tensor decomposition where the dependence
on the polarization vectors (or spin) is captured by a basis of rank-two
tensors, and the coefficients (called form-factors) can be calculated
numerically. We use the same auxiliary tensor basis $ \big\{ T_1^{\alpha\beta},
\ldots, T_9^{\alpha\beta} \big\}$ as defined in
ref.~\cite{Bohnenblust:2023qmy}. Those we contract with the polarization
vectors for $i=1,2$ defined through boost from the $\bar p'_i$ frame
\begin{align}
T_{n,i}^{v'v} = \bar\varepsilon_{v'}(p'_i)\cdot T_n \cdot \varepsilon_{v}(p_i) 
	= \bar\varepsilon_{v'}(\bar{p}'_i)\cdot\left[\Lambda(\bar{p}'_i,p'_i)\cdot T_n \cdot \Lambda(p_i,\bar{p}'_i) \right]\cdot \varepsilon_{v}(\bar{p}'_i) \,.
\end{align}
We relate them to the classical spin and define
\begin{align}\label{eqn:clRepl}
T_{n,i}^{\rm cl}(\cS) = -\big[\Lambda(\bar{p}'_i,p'_i)\cdot T_{n,i} \cdot \Lambda(p_i,\bar{p}'_i) \big]_{\mu\nu}
	\left[ \frac{1}{3}\left(\eta^{\mu\nu}-\frac{\bar{p}'^{\mu}_i \bar{p}'^{\nu}_i}{m_i^2} \right) 
	-\frac{i}{2m_i}\epsilon^{\mu \nu \rho \sigma}\bar{p}'_{i\rho} \cS_{i\sigma}
	+\cS_i^{\{\mu} \cS_i^{\nu\}}  \right]\,.
\end{align}
Finally, we reconstruct the analytic dependence on $x$ from multiple numerical
evaluations, which allows us to take the classical limit $x\to \infty$.  The so
obtained classical scattering amplitude can be directly compared to the
three-body amplitude computed within the WQFT formalism.

%--------------------------------------------------------------------------------
\section{Results}
\label{sec:results}
%-------------------------------------------------------------------------------
In this section, we collect the classical momentum-space waveform, excluding the 
(not detectable) static background contribution, up to $\mathcal{O}(G^3 \cS^2)$.
We split the discussion into the three-body waveform and the two-body waveform.
We further discuss the
regularization of its divergences and other analytic properties.

%-------------------------------------------------------------------------------
\subsection{Leading-order three-body momentum-space waveform}
%-------------------------------------------------------------------------------
As explained in detail in \cref{sec:computation}, we compute as a side product
the leading-order waveform for the scattering of three spinning black holes including 
quadratic spin terms,
\begin{align}
   \mathcal{M}_s^{3,\mathrm{LO}} = {\cal M}_s^{3}(1,2,3).
\end{align}
The waveform is obtained from $3\to 4$ tree-level
amplitudes and has the following multipole expansion
\begin{align} \label{eq:MultipoleExpansionSpin}
	\mathcal{M}_s^{3,\mathrm{LO}} = t_0 + \sum_{i=1}^3 t_{\mu\nu}^{(i)} \cS_i^{\mu\nu} 
	+ \sum_{i,j=1}^3 t_{\mu\nu\rho\sigma}^{(ij)} \cS_i^{\mu \nu}\cS_{j}^{\rho \sigma}
\end{align}
in terms of the anti-symmetric coefficients $t_{\mu\nu}^{(i)}=t_{[\mu\nu]}^{(i)}$ 
and $t_{\mu\nu\rho\sigma}^{(ij)}= t_{[\mu\nu][\rho\sigma]}^{(ij)}$.
We are providing the analytic results for these multipole coefficients in the
ancillary files of this article. We would like to emphasize the fact that the result
is expressed in terms of $14$ kinematic invariants, see \cref{sec:notation},
which are linear independent for general $d$-dimensional external kinematics.
The number of kinematic variables can be further reduced for strictly
four-dimensional momenta using Gram-determinant relations and might lead
to more compact expressions.

%-------------------------------------------------------------------------------
\subsection{Next-to-leading order two-body momentum-space waveform}
%-------------------------------------------------------------------------------
For the scattering of two black holes, we give the waveform up to NLO in $G$
and quadratic in spin, i.e at $\mathcal{O}(G^3\cS^2)$. This captures several
spin-interaction effects of the scattering. At
linear order in spin, we obtain the spin-orbit coupling $v_i \cdot \cS_j$, while at
quadratic order in spin, we additionally capture the spin-spin couplings
$\cS_1\cdot \cS_2$, and the first self-interaction spin
contributions $\cS_i^2$.

To exclude the static background, we define the momentum space LO and NLO binary waveform through
\begin{align}
	 \mathcal{M}^{2,\mathrm{LO}}_{s} &=  \cM_{\mathrm{tree}} =  \mathcal{M}_s^2(1,2)  \,, \\
	\mathcal{M}^{2,\mathrm{NLO}}_{s} &= \mathcal{M}_s^2(1,2) + \mathcal{M}_s^3(1,1,2)  + \mathcal{M}_s^3(1,2,2)\,,
\end{align}
where we introduced the redundant notation $\cM_{\mathrm{tree}}$ for convenience
and consistency with ref.~\cite{Bohnenblust:2023qmy}.
We split the amplitude into its tree-level contribution, an infrared (IR)
divergent piece, an ultraviolet (UV) divergent piece, a scale dependent tail
contribution and a finite contribution. The waveform thus reads
\begin{align}
\mathcal{M}^{2,\mathrm{NLO}}_{s} = \cM_{\mathrm{\mathrm{tree}}}  + \cM_{\mathrm{IR}}+\cM_{\mathrm{UV}}+\cM_{\mathrm{tail}}+\cM_{\mathrm{finite}}\,.
\end{align}
Next, we briefly discuss each contribution to the NLO waveform.
%-------------------------------------------------------------------------------
\paragraph{Leading-order waveform:}
%-------------------------------------------------------------------------------
The leading-order piece of the two-body waveform up to $\cS^2$ was first calculated in refs.~\cite{Jakobsen:2021lvp} 
(later confirmed by refs.~\cite{DeAngelis:2023lvf, Brandhuber:2023hhl, Aoude:2023dui}) and we find an equivalent expression.
The five-point tree amplitude contains terms suppressed by the dimensional regulator $\epsilon$. 
While not relevant at leading order, we keep it in the IR subtraction at higher order. 
We split the tree as
\begin{align}
    \cM_{\mathrm{tree}} = \cM_{\mathrm{tree}}^{\epsilon^0} + \epsilon\, \cM_{\mathrm{tree}}^{\epsilon^1} + \mathcal{O}(\epsilon^2)\,.
\end{align}
where the higher-order terms in $\epsilon$ arise working in the 't Hooft-Veltman scheme. Namely, we
performed the Lorentz algebra in $d$ dimensions, while keeping the external states in four dimensions.
The term $\cM_{\mathrm{tree}}^{\epsilon^1}$ becomes relevant in our definition of the NLO waveform.
%-------------------------------------------------------------------------------
\paragraph{Infrared divergence:}
%-------------------------------------------------------------------------------
As in the scattering-amplitude approach, the gravitational waveform
develops IR divergences.  The IR divergences factorize and their
contribution takes the form
\begin{align}
    \cM_{\mathrm{IR}} = \left[ \frac{1}{\epsilon} - \log\left(\frac{\mu^2_{\rm IR}}{\mu^2}\right)\right] \mathcal{W}_S\, \cM_{\mathrm{tree}}\;,
\end{align}
where $\mathcal{W}_S$ is the soft factor of the waveform \cite{Weinberg:1965nx,
Brandhuber:2023hhy, Bohnenblust:2023qmy} and is given by
\begin{align}
\mathcal{W}_S =  -i G (m_1 \omega_1+  m_2 \omega_2) 
    \left[ 1 + \frac{y(2y^2-3)}{2(y^2-1)^{3/2}}\right]\;.
\end{align}
As expected, the IR divergence scales with the tree. Because of this, we can
exponentiate it. To obtain a finite waveform, we may absorb this IR divergence into
a phase \cite{Goldberger:2009qd,Porto:2012as,Herderschee:2023fxh,Brandhuber:2024lgl}. 
The waveform in time domain is computed through a Fourier
transformation from frequency domain to time domain, given in
\cref{eq:DefinitionWaveform}. This Fourier transformation includes
a phase depending on the retarded-time variable, which is used to absorb the IR divergence:
\begin{align}
    e^{-i \omega (t-|{\bf r}|)} \mathcal{M}^{2,\mathrm{NLO}}_{s}  = e^{-i\omega\Big[t-|{\bf r}|+\left(\frac{1}{\epsilon}-\log \frac{\mu_{\mathrm{IR}^2}}{\mu^2} \right)\frac{\mathcal{W}_S}{i \omega} \Big]}\Big(& \cM_{\mathrm{tree}} +\cM_{\mathrm{UV}}+\cM_{\mathrm{tail}}\nn \\
    &+\cM_{\mathrm{finite}} \Big) + \mathcal{O}(G^3)\,.
\end{align}
Note that we have absorbed a term  $\mathcal{W}_S \cM_{\mathrm{tree}}^{\epsilon^1}$ into the finite terms
after IR subtraction. This piece is related to a
$\epsilon/\epsilon$ contribution from the IR. 

It is interesting to note that the redefinition of retarded time $t-|{\bf r}|$
may also be found from a general relativistic approach
\cite{Caron-Huot:2023vxl,Bohnenblust:2023qmy}. It is related to the Shapiro
delay of a graviton climbing out of a potential well to infinity
\cite{Shapiro:1964uw} and the deflection of the incoming trajectory of the two
scattering black holes due to the Kerr spacetime at infinity.

%-------------------------------------------------------------------------------
\paragraph{Ultraviolet divergences:}
%-------------------------------------------------------------------------------
As in the scattering amplitude approach 
\cite{Herderschee:2023fxh, Brandhuber:2023hhy, Georgoudis:2023lgf,
Elkhidir:2023dco,Bohnenblust:2023qmy}, the two-body waveform develops
an UV divergence in the WQFT as well. Its contribution takes the form
\begin{align}
    \cM_{\mathrm{UV}} =  \left[ \frac{1}{\epsilon} - \log\left(\frac{\mu^2_{\rm UV}}{\mu^2}\right)\right] \bar{\cM}_{\mathrm{UV}}\,.
\end{align}
Here, $\bar{\cM}_{\mathrm{UV}}$ depends polynomially on the momentum exchanges
$q_1$ and $q_2$. After a Fourier transformation from momentum to impact
parameter space, they will only yield contributions local in the impact
parameter $b$ (i.e. $\sim\delta(|b|$). We may thus drop them when considering
the far-field waveform.

%-------------------------------------------------------------------------------
\paragraph{Tail contribution:}
%-------------------------------------------------------------------------------
The tail contribution of the waveform amounts to 
\begin{align}
    \cM_{\mathrm{tail}} = - \log \left(\frac{\omega_1 \omega_2}{\mu_{\mathrm{IR}}^2}\right) \mathcal{W}_S \cM_{\mathrm{tree}} - \log  \left(\frac{\omega_1 \omega_2}{\mu_{\mathrm{UV}}^2}\right) \bar{\cM}_{\mathrm{UV}}\,. 
\end{align}
While the UV pole including the factor $\bar{\cM}_{\mathrm{UV}}$ does not contribute
to the far-field waveform, the tail has been found to dominate in the scattering waveform of
two Schwarzschild black holes \cite{Herderschee:2023fxh}.

%-------------------------------------------------------------------------------
\paragraph{Finite contributions:}
%-------------------------------------------------------------------------------
At last, the finite remainder of the NLO waveform can be written as a linear
combination of special functions $f_i$ that span the linear independent
functions of all necessary Feynman integrals.  We can write the finite
remainder as
\begin{equation}
 \cM_{\mathrm{finite}} = \sum_{i=1}^{18} r_i\,f_i\;,
\end{equation}
where the functions $f_i$ can be found in the appendix of
ref.~\cite{Bohnenblust:2023qmy} (after adjusting to the different convention by 
mapping $\omega_i \rightarrow - \omega_i$). We provide the coefficients $r_i$ 
and the definition of the functions $f_i$ in the ancillary files.

%-------------------------------------------------------------------------------
\paragraph{Ancillary files:}
%-------------------------------------------------------------------------------
We provide ancillary files containing the two-body and three-body waveform in 
momentum space up to $\mathcal{O}(G^3 \cS^2)$. Their folder structure is:
\begin{enumerate}
	\item[] \texttt{anc/waveform.m}
	\item[] \texttt{anc/waveform3BH.m}
	\item[] \texttt{anc/loadWaveform.wl}
\end{enumerate}
The file \texttt{waveform.m} contains the different contributions to the
two-body waveform up to quadratic order in spin. The spin dependence is given
in terms of the spin tensors $\cS_i^{\mu \nu}$ contracted with external
momenta, other spin tensors or itself. The waveform has been contracted with a
gravitational-wave polarization tensor $\varepsilon^{\mu\nu}(k) = \varepsilon^{\mu}(k)
\varepsilon^{\nu}(k)$.  The analytic expression for the three-body waveform can
be found in the file \texttt{waveform3BH.m}. To access the waveform, the file
\texttt{loadWaveform.wl} contains commands to load and use the results.  It
furthermore contains details on the notation.

%--------------------------------------------------------------------------------
\subsection{BMS frame}\label{sec:BMS}
%--------------------------------------------------------------------------------

By ignoring contributions due to zero-frequency modes of the gravitational wave
($\sim\delta(\omega)$), we have automatically chosen the \textit{canonical}
Bondi-Metzner-Sachs (BMS) frame
\cite{Bondi:1962px,Sachs:1962wk,Veneziano:2022zwh}. The methods traditionally
used to calculate gravitational waves emitted during the inspiral and merger of
two black holes are on the other hand in the \textit{intrinsic} BMS frame
\cite{Georgoudis:2023eke,Bini:2024rsy}. The results in the two frames are related by a
supertranslation and one can recover the waveform in the intrinsic frame form
the waveform in the canonical frame by shifting the time $t$
\cite{Georgoudis:2023eke,Georgoudis:2024pdz,Elkhidir:2024izo}
\begin{align}
 t \rightarrow t + 2 G\left(m_1 v_1\cdot \hat{k} \log(v_1\cdot \hat{k})+m_2 v_2\cdot \hat{k} \log(v_2\cdot \hat{k}) \right)\;,
\end{align}
where $\hat{k} = k/\omega$. This shift also recovers the static background at $\mathcal{O}(G)$ \cite{Veneziano:2022zwh}.

%--------------------------------------------------------------------------------
\subsection{Validation}

We validate our results through analytic consistency checks and numerical
comparisons at various stages of the computation.
A discussion of the behavior of the waveform under supersymmetry transformations
can be found in~\cref{app:susy}.

\paragraph{Gauge invariance:} We explicitly verify the gauge invariance of both
the three-body and two-body waveforms by confirming that 
\begin{align}
\lim_{k^2\rightarrow 0} \, k^2 k_\mu h_{\rm cl}^{\mu \nu}(k)  = 0
\end{align}
in both cases.

\paragraph{Three-body waveform:}  The scalar component of our three-body
waveform reproduces the result of ref.~\cite{Jakobsen:2021smu}. In addition, we
perform a numerical cross-check against the independent QFT computation
including all spin-squared contributions, as discussed in \cref{sec:QFT}. 

\paragraph{WQFT integrals:}  We validate the WQFT integrals against 
refs.\cite{Brandhuber:2023hhy,Herderschee:2023fxh,Caron-Huot:2023vxl,Bohnenblust:2023qmy} 
and by numerically evaluating them using \textsc{AMFlow}~\cite{Liu:2022chg} and find agreement with our
analytic expressions.

\paragraph{Two-body waveform:}  For the two-body case, we benchmark the
spinless NLO waveform against previous results~\cite{Brandhuber:2023hhy,
Herderschee:2023fxh, Georgoudis:2023lgf, Georgoudis:2023ozp,
Bohnenblust:2023qmy}, and confirm agreement. The linear-in-spin contribution is
likewise consistent with ref.~\cite{Bohnenblust:2023qmy}. Furthermore, we
verify that the IR divergence matches the expectation from Weinberg's 
soft-graviton theorem, while the UV divergence does not contribute to the final observable
waveform. Moreover, the two-body waveform is constructed from the previously validated
three-body waveform. Finally, we verify that the NLO waveform is regular in phase space, 
by confirming that spurious poles in the function coefficients cancel.
%--------------------------------------------------------------------------------
\section{Conclusion}
\label{sec:conclusions}
%--------------------------------------------------------------------------------
In this work, we presented the quadratic-in-spin contributions to the gravitational 
waveform emitted during
the scattering of two spinning black holes at next-to-leading PM order, i.e at
$\mathcal{O}(G^3\cS^2)$. 
Our calculation is based on the WQFT formalism~\cite{Mogull:2020sak,Kalin:2020mvi,Jakobsen:2022psy,Kalin:2022hph}
that allows to compute
classical observables directly from tree-level amplitudes in the
Keldysh-Schwinger formalism \cite{Schwinger:1960qe,Keldysh:1964ud}. In particular, 
we closely follow the supersymmetric variant
\cite{Jakobsen:2021lvp,Jakobsen:2021zvh} of the WQFT formalism to obtain
classical spin effects. % to quadratic order.
We have extensively reviewed the formalism and
showed the expected consistency \cite{Damgaard:2023vnx} between the WQFT and scattering-amplitude results.
In our discussion, we placed particular emphasis on
identifying generating WQFT amplitudes (see e.g.~\cite{Mogull:2020sak}), 
which resemble loop-amplitude integrands 
and may equivalently be obtained from tree-level amplitudes in field-theory \cite{Shen:2018ebu}. 
Furthermore, we confirmed the applicability of the cut-improved 
Feynman integral basis
\cite{Caron-Huot:2023vxl} for the retarded radiation dynamic. 
As a by-product,
we also obtained for the first time the LO gravitational
waveform for the scattering of three spinning black holes.

To further highlight the similarities between the two approaches, we verified
our WQFT results 
by comparing to scattering amplitudes of scalars, Proca fields and gravitons.
The latter approach uses
familiar Feynman rules and provides a sturdy cross‑check, but obtaining
classical observables though loop computations appeared to be less direct: hyper‑classical terms appear and cancel
only in the final result \cite{ Herderschee:2023fxh, Brandhuber:2023hhy, Georgoudis:2023lgf,
Elkhidir:2023dco, Bohnenblust:2023qmy}. 
Furthermore, re‑expressing polarization vectors in terms of
classical spin—while keeping the correct Casimir contribution—can be subtle.
Performing these steps numerically and reconstructing only the newly introduced generating 
%seven-point tree-level 
amplitude proved useful. 
In this comparison we extracted the spin Casimir operator
contributions by employing the spin-interpolation method~\cite{Akpinar:2024meg,
Akpinar:2025bkt} which adds further evidence for its validity.  Taken together,
the WQFT offers a direct route to analytic expressions, while the scatting-amplitudes approach supplies a robust
numerical validation, and we exploit the strengths of both frameworks
throughout this work.

We performed several additional validation steps on our results, such as checking Ward
identities, consistency with the known factorization of infrared
divergences~\cite{Weinberg:1965nx}, and reproducing the known spin-independent 
\cite{Brandhuber:2023hhl,
DeAngelis:2023lvf, Aoude:2023dui,Brandhuber:2024qdn,Bohnenblust:2023qmy}
and
linear-in-spin contributions to the NLO waveform~\cite{Bohnenblust:2023qmy}.
Finally, we confirmed the expected regularity of the waveform in phase space.

One natural follow-up direction is to extend our current techniques by
employing the recent higher-spin extension of the WQFT
formalism~\cite{Haddad:2024ebn} to obtain the gravitational waveform up to the
quartic order in spin. To obtain the actual observable analytically, it remains
to perform the  Fourier transformation to impact-parameter space.  This has
turned out to be a highly non-trivial task~\cite{Herderschee:2023fxh,
Bohnenblust:2023qmy} 
%and we expect this task to be more challenging for
and we expect further challenges in
higher-spin contributions. However, we anticipate that the method of
ref.~\cite{Brunello:2024ibk} will simplify this step tremendously.  Additionally,
so far our results describe scattering of black holes.  It
would be interesting to make use of the analytic continuation of the waveform,
as proposed in refs.~\cite{Adamo:2022ooq, Adamo:2024oxy}, to obtain a
gravitational waveform also for bound binary systems of spinning black holes.
Finally, our results can be used to obtain the angular momentum and
radiation reaction at $\mathcal{O}(G^4)$~\cite{Heissenberg:2025ocy} for
rotating black holes.  We leave these open questions for future investigation.

%--------------------------------------------------------------------------------
\section*{Acknowledgments}
%--------------------------------------------------------------------------------
L.B. acknowledges support from the Swiss National Science Foundation (SNSF) under
the grant 200020 192092. L.B. thanks the UZH Candoc Grant for financial support. 
M.K. is supported by the DGAPA-PAPIIT grant IA102224
(“Iluminando agujeros negros”) at UNAM.  

\appendix
%--------------------------------------------------------------------------------
\section{Feynman rules}
\label{app:feynmanrules}
%--------------------------------------------------------------------------------
In this appendix we provide the vertex Feynman rules extracted from
the WQFT action~\eqref{eq:FullAction}.  For graviton
self-interactions we make use of the \textsc{FeynGrav}
program~\cite{Latosh:2023zsi} to compute the corresponding Feynman rules.

In the following we will often need to symmetrize in Lorentz indices, and we introduce 
the operator $P^{(\mu\nu)}$ which symmetrises tensors $t_{\mu\nu}$ 
\begin{align}
P^{(\mu_1,\nu_1)} ( t_{\mu_1\nu_1} ) = t_{(\mu_1\nu_1)}\,,
\end{align}
following the notation of \cref{eqn:tensorSym}.
For consecutive symmetrization we use the notation 
\begin{align}
P^{(\mu_1,\nu_1)\ldots (\mu_n,\nu_n)} = P^{(\mu_1,\nu_1)} \cdots  P^{(\mu_n,\nu_n)} \,.
\end{align}

%--------------------------------------------------------------------------------

%--------------------------------------------------------------------------------
\subsubsection*{One-graviton worldline vertex}
%--------------------------------------------------------------------------------
\begin{align}   
    \raisebox{-25pt}{\includegraphics[]{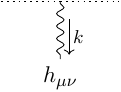} } =
  -i\frac{m\kappa }{2}e^{ik\cdot b}\dh(k\cdot v)\Big[
  v^{(\mu} v^{\nu)}
  +\frac{i}{m} (k\cdot\cS)^{(\mu}v^{\nu)}-\frac{1}{2 m^2}(k\cdot \cS)^{(\mu} (k\cdot \cS)^{\nu)}\Big]
  \label{eq:h}
\end{align}

%--------------------------------------------------------------------------------
\subsubsection*{One-graviton one-scalar worldline vertex}
%--------------------------------------------------------------------------------
\begin{align}
    \raisebox{-25pt}{\includegraphics[]{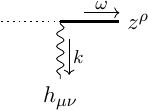}} &=
  \frac{m\kappa }{2}e^{ik\cdot b}\dh(k\cdot v+\omega)
  \bigg( 2\omega v^{(\mu}\delta^{\nu)}_\rho+v^{(\mu} v^{\nu)} k_\rho
  +\frac{i}{m} (k\cdot\cS)^{(\mu}(k_\rho v^{\nu)}+\omega\delta_\rho^{\nu)})\nn \\[-5mm]
  &+\frac{1}{2 m^2}(k\cdot \cS)^{\mu}(\cS\cdot k)^{\nu}k_{\rho}  \bigg)
  \label{eq:hz}
\end{align}

%--------------------------------------------------------------------------------
\subsubsection*{One-graviton one-Grassmann-field vertex}
%--------------------------------------------------------------------------------
\begin{align}
    \raisebox{-25pt}{\includegraphics[]{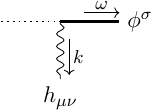}} = -im\kappa e^{ik\cdot b}\dh(k\cdot v+\omega) 
    \bigg(k_{[\sigma}\delta_{\xi]}^{(\mu}\!v^{\nu)} - \frac{i}{m} k_{[\sigma}\delta_{\xi]}^{(\mu} (\cS \cdot k)^{\nu )}
      \bigg)\bar{\Psi}^\xi
     \label{eq:hPhi}
\end{align}

%--------------------------------------------------------------------------------
\subsubsection*{One-graviton two-scalar worldline vertex}
%--------------------------------------------------------------------------------
\begin{align}
    \raisebox{-40pt}{\includegraphics[]{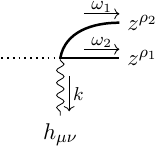}}
&=
m \kappa e^{ik\cdot b}\dh(k\cdot v+\omega_1+\omega_2)\times\nn\\[-10mm]
&\Bigg[i\bigg(\frac{1}{2}k_{\rho_1} k_{\rho_2}v^{(\mu} v^{\nu)}+
\omega_1k_{\rho_2} v^{(\mu}\delta^{\nu)}_{\rho_1}+
\omega_2k_{\rho_1} v^{(\mu}\delta^{\nu)}_{\rho_2}+
\omega_1\omega_2\delta^{(\mu}_{\rho_1}\delta^{\nu)}_{\rho_2}\bigg)\nn\\
&+\frac{1}{2 m} (\cS\cdot k_{\sigma})^{(\mu}\big(\omega_1 k_{\rho_2} \delta^{\nu)}_{\rho_1}+\omega_2 k_{\rho_1} \delta^{\nu)}_{\rho_2} + k_{\rho_1}k_{\rho_2} v^{\nu)} \big)\nn\\
&-\frac{1}{4 m^2}i (\cS\cdot k)^{(\mu}(\cS\cdot k)^{\nu)}k_{\rho_1}k_{\rho_2}
\Bigg]
\end{align}

%--------------------------------------------------------------------------------
\subsubsection*{One-graviton one-scalar one-Grassmann-field worldline vertex}
%--------------------------------------------------------------------------------
\begin{align}
    \raisebox{-40pt}{\includegraphics[]{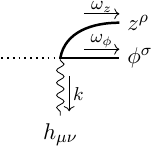}}
&=
-\frac{m\kappa}{2} e^{ik\cdot b}\dh(k\cdot v+\omega_z+\omega_\phi) \bigg[ \Big( \delta^{(\mu}_{\sigma}(k\cdot \bar{\Psi}) - k_{\sigma}\bar{\Psi}^{(\mu} \Big) \nn \\[-10mm] 
&\times \Big(v^{\nu)}k_{\rho} + \omega_z \delta^{\nu)}_{\rho} - \frac{i}{m} (\cS \cdot k)^{\nu)} k_{\rho} \Big) 
\bigg]
\label{eq:hzPhi}
\end{align}

%--------------------------------------------------------------------------------
\subsubsection*{One-graviton two-Grassmann-field worldline vertex}
%--------------------------------------------------------------------------------
\begin{align}
    \raisebox{-40pt}{\includegraphics[]{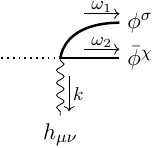}}
&=i m \kappa e^{ik\cdot b}\dh(k\cdot v+\omega_1+\omega_2)\bigg(  \delta_{[\sigma}^{(\mu}k_{1,\chi]}v^{\nu)}-\frac{1}{2} \delta_{\chi}^{(\mu}\delta_{\sigma}^{\nu)}(k\cdot\Psi)(k\cdot\bar{\Psi})\nn \\[-10mm]
 &+\frac{1}{2} k_{\chi} \delta_{\sigma}^{(\mu} (k\cdot \Psi) \bar{\Psi}^{\nu)} - \frac{1}{2} k_{\sigma} \delta_{\chi}^{(\mu}(k\cdot \bar{\Psi}) \Psi^{\nu)}-\frac{1}{2} k_{\sigma}k_{\chi}\Psi^{(\mu}\bar{\Psi}^{\nu )}\bigg)
\label{eq:hPhiPhiBar}
\end{align}

\begin{align}
    \raisebox{-40pt}{\includegraphics[]{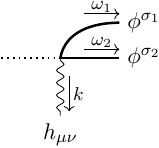}}
&=i m \kappa e^{ik\cdot b}\dh(k\cdot v+\omega_1+\omega_2) \delta^{(\mu}_{[\sigma_1}k_{\sigma_2]}\bar{\Psi}^{\nu)} (k\cdot \bar{\Psi})
\label{eq:hPhiPhi}
\end{align}

%--------------------------------------------------------------------------------
\subsubsection*{Two-graviton worldline vertex}
%--------------------------------------------------------------------------------
\begin{align}
    \raisebox{-25pt}{\includegraphics[]{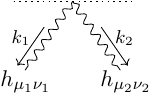}}
=& P^{(\mu_1,\nu_1)(\mu_2,\nu_2)}\Big[ -\frac{\kappa^2}{4} e^{i(k_1+k_2)\cdot b}\dh((k_1+k_2)\cdot v) \bigg( (k_1\cdot S)^{\mu_2} v^{\mu_1} \eta^{\nu_1 \nu_2}  \label{eq:hh} \\[-6mm]
& - S^{\mu_1 \mu_2} (v^{\nu_1}k_1^{\nu_2}-\frac{1}{2} k_1\cdot v \eta ^{\nu_1 \nu_2})+\frac{i}{m} \bigg[\Big((S\cdot k_1)^{\mu_1} (S \cdot k)^{\mu_2}\nn\\
 &+ \frac{1}{2}(S\cdot k_2)^{\mu_1}(S\cdot k_1)^{\mu_2} - \frac{1}{2} S^{\mu_1 \mu_2} (k_1\cdot S \cdot k_2)  \Big)\eta^{\nu_1 \nu_2} \nn \\
 &+\frac{1}{4} k_1\cdot k_2 S^{\mu_1 \nu_2}S^{\mu_2 \nu_1} - k_1^{\nu_2} (S\cdot(k_1+k_2)^{\mu_1})S^{\mu_2 \nu_1} \bigg] 
   \bigg) + (1 \leftrightarrow 2) \Big] \nn
\end{align}
%which is symmetrized over all $(\mu_i,\nu_i)$-pairs. 

%--------------------------------------------------------------------------------
\subsubsection*{Two-graviton one-scalar worldline vertex}
%--------------------------------------------------------------------------------
\begin{align}
    \raisebox{-25pt}{\includegraphics[]{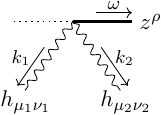}}
=&
 P^{(\mu_1,\nu_1)(\mu_2,\nu_2)}\Big[ 
 \frac{\kappa^2}{4} e^{i(k_1+k_2)\cdot b}\dh((k_1+k_2)\cdot v+\omega)  \bigg( i (k_1+k_2)_{\rho}\big[v^{\mu_1}k_1^{\mu_2} S^{\nu_1\nu_2} 
\nn\\[-6mm] 
&+\eta^{\mu_1\mu_2} (S\cdot k_1)^{\nu_2} v^{\nu_1}-\frac{1}{2}\eta^{\mu_1 \mu_2} k_1\cdot v S^{\nu_1\nu_2} \big]+i \omega \big[\delta^{\mu_1}_{\rho} \eta^{\nu_1\nu_2} (S\cdot k_1)^{\mu_2}\nn \\
&+ \delta^{\mu_1}_{\rho} k_1^{\mu_2} S^{\nu_1\nu_2} - \frac{1}{2} \eta^{\mu_1 \mu_2} k_{1\rho} S^{\nu_1 \nu_2} \big] 
\label{eq:hhz} 
\\
&+\frac{1}{m}(k_1+k_2)_{\rho}\big[\eta^{\mu_1\mu_2}(S\cdot k_1)^{\nu_2} + k_1^{\mu_2} S^{\mu_1 \nu_2} \big](S\cdot k_1)^{\nu_1}\bigg)+ (1 \leftrightarrow 2) \Big] \nn
\end{align}
%which is symmetrized over all $(\mu_i,\nu_i)$-pairs.
%--------------------------------------------------------------------------------
\subsubsection*{Two-graviton one-Grassmann-field worldline vertex}
%--------------------------------------------------------------------------------
\begin{align}
    \raisebox{-25pt}{\includegraphics[]{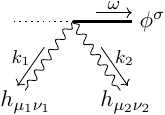}}
&=
 P^{(\mu_1,\nu_1)(\mu_2,\nu_2)}\Big[ 
\frac{m \kappa^2}{4} e^{i(k_1+k_2)\cdot b}\dh((k_1+k_2)\cdot v+\omega) \bigg( i\big[\eta^{\mu_1 \mu_2} k_{1 \sigma} v^{\nu_1} \bar{\Psi}^{\nu_2}\nn\\[-6mm] &+\delta^{\mu_1}_{\sigma}\big(k_2^{\nu_1}v^{\mu_2}-k_1^{\mu_2}v^{\nu_1}+\frac{1}{2}(k_1-k_2)\cdot v \eta^{\mu_2 \nu_1}  \big)\bar{\Psi}^{\nu_2}-\eta^{\mu_1 \mu_2}\delta^{\nu_2}_{\sigma} v^{\nu_1} k_1\cdot \bar{\Psi}    \big]  \bigg) \nn\\
&+\frac{1}{m} \big[\eta^{\mu_1 \mu_2}\big((S\cdot k_1)^{\nu_1} k_{1 \sigma}+ (S\cdot k_2)^{\nu_1} k_{2 \sigma}   \big) \bar{\Psi}^{\nu_2} + \delta^{\mu_1}_{\sigma}\big( (S\cdot k_2)^{\mu_2} k_2^{\nu_1} \nn\\
 &- (S\cdot k_1)^{\nu_1}k_1^{\mu_2}  \big)\bar{\Psi}^{\nu_2} - \delta^{\nu_1}_{\sigma}k_1^{\nu_2}S^{\mu_1\mu_2} k_1 \cdot \bar{\Psi}-\delta^{\mu_1}_{\sigma} \eta^{\nu_1\mu_2} \big(k_1\cdot \bar{\Psi} (S\cdot k_1)^{\nu_2} \nn\\
 &+ k_2\cdot \bar{\Psi} ( S \cdot k_2)^{\nu_2} \big)-k_{2\sigma}k_2^{\mu_1}S^{\nu_1 \mu_2} \bar{\Psi}^{\nu_2}\big]+ (1 \leftrightarrow 2)  \Big] \label{eq:hhPhi} 
\end{align}
%which is symmetrized over all $(\mu_i,\nu_i)$-pairs.
%--------------------------------------------------------------------------------
\subsubsection*{Three-graviton worldline vertex}
%--------------------------------------------------------------------------------

\begin{align}
    \raisebox{-25pt}{\includegraphics[]{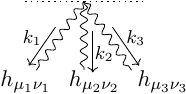}}
=&
 P^{(\mu_1,\nu_1)(\mu_2,\nu_2)(\mu_3,\nu_3)}\Big[ 
 \frac{ \kappa^3}{8} e^{i(k_1+k_2+k_3)\cdot b}\dh((k_1+k_2+k_3)\cdot v)\label{eq:hhh}  \\[-6mm] 
	& \times\bigg(g^{\mu_2 \mu_3}\Big[(g^{\mu_1 \nu_3}k_1\cdot v  
+ k_3^{\mu_1}v^{\nu_3})S^{\nu_1 \nu_2} + \frac{3}{2}(k_1^{\nu_3} S^{\nu_2 \mu_1} - g^{\mu_1 \nu_3}(S\cdot k_1)^{\nu_2})v^{\nu_1} \Big]\nn \\
&+\frac{i}{m} \Big[ g^{\mu_2 \mu_3} g^{\mu_1 \nu_3}(S\cdot(k_2 + \frac{3}{2}k_1))^{\nu_1}(S\cdot k_1)^{\nu_2} \nn\\
&+ \frac{1}{2} g^{\mu_1 \mu_2} g^{\mu_3 \nu_1} (S\cdot k_1)^{\nu_2} (S\cdot k_1)^{\nu_3} -g^{\mu_2 \mu_3}\big[(S\cdot k_2)^{\nu_3} k_2^{\mu_1} \nn \\
&- 2 (S\cdot k_2)^{\mu_1}k_1^{\nu_3}+k_3^{\mu_1}(S\cdot k_3)^{\nu_3}+g^{\mu_1 \nu_3} k_1\cdot S \cdot k_2 \big] S^{\nu_1 \nu_2}   \nn \\
&+ \frac{1}{2}\big[k_2^{\mu_1} k_2^{\nu_3} S^{\mu_3 \mu_2}-k_1^{\nu_3} k_2^{\mu_3} S^{\mu_1 \mu_2}-2g^{\mu_1\mu_3} (S\cdot k_2)^{\mu_2} k_2^{\nu_3} \big] S^{\nu_1 \nu_2} \Big]      \bigg)  \Big] \nn \\
&+ \mathrm{permutations}(1,2,3) \nn
\end{align}
%which is symmetrized over all $(\mu_i,\nu_i)$-pairs.

\section{Worldline supersymmetry property}
\label{app:susy}

We confirm that the two-body
and three-body integrand both are consistent with eq.~(26) of
ref.~\cite{Jakobsen:2021lvp}, that links higher-spin orders to lower-spin. 
These relations can be derived from the worldline-supersymmetry transformations
and are naturally fulfilled when spin tensors obey the covariant
spin-supplementary condition \eqref{eqn:ssc}.
In fact, in
the three-body case, the relations are
\begin{align}
	\frac{i}{2}q_{i\mu} t_0 = v_{i}^{\nu} t_{\mu \nu}^{(i)}\,, \qquad 
	\frac{i}{4}q_{j\rho} t_{\mu \nu}^{(i)} = v_{j}^{\sigma} t_{\mu\nu \rho \sigma}^{(i j)} \,,
\end{align}
where the tensors $t_0$, $ t_{\mu \nu}^{(i)}$, and $t_{\mu\nu \rho
\sigma}^{(i j)}$ have been introduced in \cref{eq:MultipoleExpansionSpin} as the
multipole coefficients of the spin expansion. These tensors are defined up to
contributions in direction of $v_i$, since they drop out once contracted with
the spin tensor \eqref{eqn:ssc}. 
We can thus ensure
the above relations by recasting the coefficients into the form
\begin{align}
	t_{\mu \nu}^{(i)} &= t_{\perp,\mu\nu}^{(i)}+i q_{i,[\mu|}v_{i,|\nu]} t_0\,,\\
%%%%%%%%%%%%%%%%%%%
	t_{\mu\nu\rho\sigma}^{(ij)}&= t_{\perp,\mu\nu\rho\sigma}^{(ij)}+
	\frac{i}{2} t_{\perp, \mu \nu}^{(i)} \, q_{j,[\rho|} v_{j,|\sigma]} +
	\frac{i}{2} q_{i,[\mu|} v_{i,|\nu]} \, t_{\perp, \rho \sigma}^{(j)} -  \\
&\quad 	\frac{1}{2} q_{i,[\mu|} v_{i,|\nu]} \, q_{j,[\rho|} v_{j,|\sigma]} \, t_0 \,.
\end{align}
Here the index $\perp$ means that a tensor does not contain any components in the
direction $v_i^{\mu}$ (or $v_j^{\mu}$) direction, i.e. $v_i^{\mu}
t_{\perp,\mu\nu}^{(i)} =v_j^{\rho}
t_{\perp,\mu\nu\rho\sigma}^{(ij)}= 0$. The expressions given in the ancillary
files are not of said form but can easily be manipulated to match it without
changing the waveform.

%--------------------------------------------------------------------------------
\bibliographystyle{JHEP}
\bibliography{main.bib}
%--------------------------------------------------------------------------------
\end{document}